\documentclass[5p,times,authoryear]{elsarticle}
\usepackage{numcompress}\biboptions{authoryear}

\usepackage{ecrc}
\usepackage{amssymb}
\usepackage{threeparttable}
\usepackage[french]{babel}
\usepackage{rotating}
\usepackage{tabularx}
\usepackage{adjustbox}
\usepackage{natbib} 
\usepackage{enumitem}

\newcommand{\Mearth}{$M_\oplus$}

\DeclareTextSymbol{\deg}{T1}{6}
\DeclareTextSymbol{\deg}{OT1}{23}

\volume{00}

\firstpage{1}

\journalname{Planetary and Space Science}

\runauth{O. Mousis et al.}

\jid{PSS}

\jnltitlelogo{}

\usepackage[bookmarks = true, bookmarksnumbered = true, pdfpagemode =None, pdfstartview = FitH, pdfpagelayout = SinglePage, colorlinks = true, urlcolor = blue, citecolor = monbleu]{hyperref}

\usepackage{amssymb}
\usepackage{lineno}

\biboptions{comma,round}

\begin{document}

\begin{frontmatter}



\title{Scientific rationale for Uranus and Neptune {\it in situ} explorations}

\author[LAM]{O. Mousis}
\ead{olivier.mousis@lam.fr}
\address[LAM]{Aix Marseille Universit{\'e}, CNRS, LAM (Laboratoire d'Astrophysique de Marseille) UMR 7326, 13388, Marseille, France
\fnref{label2}}
\author[JPL]{D.~H. Atkinson}
\address[JPL]{Jet Propulsion Laboratory, California Institute of Technology, 4800 Oak Grove Dr., Pasadena, CA 91109, USA}
\address[LESIA]{ LESIA, Observatoire de Paris, PSL Research University, CNRS, Sorbonne Universit\'es, UPMC Univ. Paris 06, Univ. Paris Diderot, Sorbonne Paris Cit\'e, 5 place Jules Janssen, 92195 Meudon, France}
\author[LESIA]{T. Cavali\'e}
\address[UL]{Department of Physics \& Astronomy, University of Leicester, University Road, Leicester, LE1 7RH, UK}
\author[UL]{L.~N. Fletcher}
\address[GSFC]{NASA Goddard Space flight Center, Greenbelt, MD 20771, USA}
\author[GSFC]{M.~J. Amato}
\author[GSFC]{S. Aslam}
\author[CISAS]{F. Ferri}
\address[CISAS]{Universit\`a degli Studi di Padova, Centro di Ateneo di Studi e Attivit\`a Spaziali ``Giuseppe Colombo'' (CISAS), via Venezia 15, 35131 Padova, Italy}
\author[LPC2E]{J.-B. Renard}
\address[LPC2E]{CNRS-Universit{\'e} d'Orl{\'e}ans, 3a Avenue de la Recherche Scientifique, 45071 Orl{\'e}ans Cedex 2, France}
\author[SSSE]{T. Spilker}
\address[SSSE]{Solar System Science \& Exploration, Monrovia, USA}
\author[Ames]{E. Venkatapathy}
\address[Ames]{NASA Ames Research Center, Moffett field, California, USA}
\address[UB]{Space Science \& Planetology, Physics Institute, University of Bern, Sidlerstrasse 5, 3012 Bern, Switzerland}
\author[UB]{P. Wurz}
\address[UOX]{Department of Physics, University of Oxford, Denys Wilkinson Building, Keble Road, Oxford OX1 3RH, UK}
\author[UOX]{K. Aplin}
\author[LESIA]{A. Coustenis}
\author[LAM]{M. Deleuil}
\author[LAB]{M. Dobrijevic}
\address[LAB]{Laboratoire d'astrophysique de Bordeaux, University Bordeaux, CNRS, B18N, all\'ee Geoffroy Saint-Hilaire, 33615 Pessac, France}
\author[LESIA]{T. Fouchet}
\author[OCA]{T. Guillot}
\address[OCA]{Observatoire de la C\^ote d'Azur, Laboratoire Lagrange, BP 4229, 06304 Nice cedex 4, France}
\address[MPS]{Max-Planck-Institut f\"ur Sonnensystemforschung, Justus von Liebig Weg 3, 37077 G\"ottingen, Germany}
\author[MPS]{P. Hartogh}
\address[UM]{University of Maryland, College Park, MD 20742, USA}
\author[UM]{T. Hewagama}
\author[JPL]{M.~D. Hofstadter}
\address[SWRI]{Southwest Research Institute, San Antonio, TX 78228, USA}
\author[SWRI]{V. Hue}
\address[UPV]{Departamento F\' isica Aplicada I, Escuela des Ingenier\'ia de Bilbao, UPV/EHU, 48013 Bilbao, Spain}
\author[UPV]{R. Hueso}
\author[LPC2E]{J.-P. Lebreton}
\author[LESIA]{E. Lellouch}
\address[SSI]{Space Science Institute, 4750 Walnut Street, Suite 205, Boulder, CO 80301, USA}
\author[SSI]{J. Moses}
\author[JPL]{G.~S. Orton}
\author[GSFC]{J.~C. Pearl}
\author[UPV]{A. S\'anchez-Lavega}
\author[GSFC]{A. Simon}
\author[LISA]{O. Venot}
\address[LISA]{Laboratoire Interuniversitaire des Syst\`emes Atmosph\'eriques (LISA), UMR CNRS 7583, Universit\'e Paris Est Cr\'eteil et Universit\'e Paris Diderot, Institut Pierre Simon Laplace, 94000 Cr\'eteil, France}
\author[SWRI]{J.~H. Waite}
\address[MICH]{Department of Atmospheric, Oceanic, and Space Sciences, University of Michigan, Ann Arbor, MI 48109-2143, USA}
\author[UM]{R.~K. Achterberg}
\author[MICH]{S. Atreya}
\author[LAB]{F. Billebaud}
\address[PIIM]{Aix-Marseille Universit\'e, PIIM UMR-CNRS 7345, F-13397 Marseille, France}
\address[IRAP]{Institut de Recherche en Astrophysique et Plan\'etologie (IRAP), CNRS/Universit\'e Paul Sabatier, 31028 Toulouse, France}
\author[IRAP]{M. Blanc}
\author[PIIM]{F. Borget}
\author[LAM]{B. Brugger}
\address[IPG]{Institut de Physique du Globe, Sorbonne Paris Cit\'e, Universit\'e Paris Diderot/CNRS, 1 rue Jussieu, 75005, Paris, France}
\author[IPG]{S. Charnoz}
\author[PIIM]{T. Chiavassa}
\author[UM]{V. Cottini}
\author[PIIM]{L. d'Hendecourt}
\author[PIIM]{G. Danger}
\author[LESIA]{T. Encrenaz}
\address[CUA]{The Catholic University of America, Washington, DC 20064, USA}
\author[CUA]{N.~J.~P. Gorius}
\author[LAM]{L. Jorda}
\address[CRPG]{CRPG-CNRS, Nancy-Universit\'e, 15 rue Notre Dame des Pauvres, 54501 Vandoeuvre-l\`es-Nancy, France}
\author[CRPG]{B. Marty}
\author[LESIA]{R. Moreno}
\address[OU]{Department of Physical Sciences, The Open University, Walton Hall, Milton Keynes MK7 6AA, UK}
\author[OU]{A. Morse}
\author[GSFC]{C. Nixon}
\author[JPL]{K. Reh}
\author[LAM]{T. Ronnet}
\author[OCA]{F.-X. Schmider}
\author[OU]{S. Sheridan}
\author[JPL]{C. Sotin}
\author[LAM]{P. Vernazza}
\author[GSFC]{G.~L. Villanueva}

\begin{abstract}
The ice giants Uranus and Neptune are the least understood class of planets in our solar system but the most frequently observed type of exoplanets. Presumed to have a small rocky core, a deep interior comprising $\sim$70\% heavy elements surrounded by a more dilute outer envelope of H$_2$ and He, Uranus and Neptune are fundamentally different from the better-explored gas giants Jupiter and Saturn. Because of the lack of dedicated exploration missions, our knowledge of the composition and atmospheric processes of these distant worlds is primarily derived from remote sensing from Earth-based observatories and space telescopes. As a result, Uranus's and Neptune's physical and atmospheric properties remain poorly constrained and their roles in the evolution of the Solar System not well understood. Exploration of an ice giant system is therefore a high-priority science objective as these systems (including the magnetosphere, satellites, rings, atmosphere, and interior) challenge our understanding of planetary formation and evolution. Here we describe the main scientific goals to be addressed by a future {\it in situ} exploration of an ice giant. An atmospheric entry probe targeting the 10-bar level, about 5 scale heights beneath the tropopause, would yield insight into two broad themes: i) the formation history of the ice giants and, in a broader extent, that of the Solar System, and ii) the processes at play in planetary atmospheres. The probe would descend under parachute to measure composition, structure, and dynamics, with data returned to Earth using a Carrier Relay Spacecraft as a relay station. In addition, possible mission concepts and partnerships are presented, and a strawman ice-giant probe payload is described. An ice-giant atmospheric probe could represent a significant ESA contribution to a future NASA ice-giant flagship mission.
\end{abstract}


\end{frontmatter}

%\linenumbers

\section{Introduction}
\label{Intro}

The ice giant planets Uranus and Neptune represent a largely unexplored class of planetary objects, which fills the gap in size between the larger gas giants and the smaller terrestrial worlds. Uranus and Neptune's great distances have made exploration challenging, being limited to flybys by the Voyager 2 mission in 1986 and 1989, respectively \citep{Lindal1987,Tyler1986,86smith,89smith,92lindal_nep,Stone1989}. Therefore, much of our knowledge of atmospheric processes on these distant worlds arises from remote sensing from Earth-based observatories and space telescopes (see e.g. \citealt{Encrenaz2000,Karkoschka2009,11karkoschka_CH$_4$,Feuchtgruber2013,10fletcher_akari,14fletcher_nep,14aorton,14borton,14sromovsky,Lellouch2015}). Such remote observations cannot provide ``ground-truth'' of direct, unambiguous measurements of the vertical atmospheric structure (temperatures and winds), composition and cloud properties.  With the exception of methane, these observations have never been able to detect the key volatile species (NH$_3$, H$_2$S, H$_2$O) thought to comprise deep ice giant clouds, and a host of additional minor species have remained hidden. Because of the physical limitations of these remote observations, and the deficiency of {\it in situ} or close-up measurements, Uranus and Neptune's physical and atmospheric properties are poorly constrained and their roles in the evolution of the Solar System are not well understood.

Uranus and Neptune are fundamentally different from the better-known gas giants Jupiter and Saturn. Interior models generally predict a small rocky core, a deep interior of $\sim$70\% of heavy elements surrounded by a more diluted outer envelope with a transition at $\sim$70$\%$ in radius for both planets \citep{Hubbard1995,Fortney2010,Helled2011}. Uranus and Neptune also have similar 16 to 17-hour rotation periods that shape their global dynamics.  For all their similarities, the two worlds are also very different. Uranus is closer to the Sun at $\sim$19 AU versus Neptune's 30 AU and the two planets receive solar fluxes of only 3.4 W/m$^2$ and 1.5 W/m$^2$, respectively. However, while Neptune has an inner heat source comparable to the heating received by the Sun, Uranus lacks any detectable internal heat \citep{91pearl}, possibly due to a more sluggish internal circulation and ice layers \citep{95smith, 17helled}.  Additionally, the two planets experience very different seasonal variations, as Uranus's  98$^\circ$ obliquity results in extreme seasons, compared with Neptune's more moderate 28$^\circ$ obliquity. These extremes of solar insolation have implications for the atmospheric temperatures, cloud formation, photochemistry and general circulation patterns.  Perhaps related to these differences, Uranus shows less cloud activity than Neptune, with infrequent storms \citep{09irwin}, while Neptune's disk was dominated by the Great Dark Spot at the time of the Voyager 2 flyby \citep{89smith,93sromovsky} and by bright cloud systems in more recent years \citep{17hueso}.

Exploration of an ice giant system is a high-priority science objective, as these systems (including the magnetosphere, satellites, rings, atmosphere, and interior) challenge our understanding of planetary formation and evolution. A mission to Uranus and Neptune could help answer why the ice giants are located at such large distances from the Sun, while several models predict their formation much closer \citep{Levison2001,Levison2008,Levison2011,Gomes2005,Morbidelli2005,Morbidelli2007,Nesvorny2011,Batygin2010,Batygin2012}. Also, $\sim$35\% of the extrasolar planets discovered to date have masses similar to those of Uranus and Neptune and are located at very different orbital distances. Hence, the {\it in situ} investigation of these planets could provide a useful context to the interpretation of exoplanet observations and favor future development of ice giant formation and evolution theories in general \citep{Schneider2011}. The importance of the ice giants is reflected in NASA's 2011 Decadal Survey, comments from ESA's Senior Survey Committee in response to L2/L3 and M3 mission proposals \citep{Arridge2012,Arridge2014,Turrini2014} and results of the 2017 NASA/ESA Ice Giants study \citep{Elliott2017}. 

Since the Voyager encounters, atmospheric processes at play in Jupiter and Saturn have been well characterized by the Galileo and Juno orbiters at Jupiter, and the Cassini orbiter at Saturn. The Galileo probe provided a step-change in our understanding of Jupiter's origins \citep{Owen1999,Gautier2001}, and similar atmospheric probes for Saturn have been proposed to build on the discoveries of the Cassini mission \citep{Spilker2011,Spilker2012,Atkinson2012,Atkinson2013,Atkinson2014,Atkinson2016,Venkatapathy2012,Mousis2014a,Mousis2016}. The cold, distant ice giants are very different worlds from Jupiter and Saturn, and remote studies are considerably more challenging and less mature.  An ice-giant probe would bring insights into two broad themes: i) the formation history of Uranus and Neptune and in a broader extent that of the Solar System, and ii) the processes at play in planetary atmospheres. The primary science objectives for an ice-giant probe would be to measure the bulk composition, and the thermal and dynamic structure of the atmosphere. The Uranus and Neptune atmospheres are primarily hydrogen and helium, with significant abundances of noble gases and isotopes that can only be measured by an {\it in situ} probe. Although the noble gases and many isotopes are expected to be well-mixed and therefore measurements in the upper atmosphere will suffice, there are also a number of condensable species that form cloud layers at depths that depend on abundance of the condensibles and the atmospheric thermal structure. Additionally, disequilibrium species upwelling from the deeper, hotter levels of Uranus and Neptune provide evidence of abundances and chemistry in deeper regions unreachable by the probe. Noble gas abundances are diagnostics of the formation conditions under which the ice and gas giants formed. The condensable species forming different cloud layers are indications of the protosolar nebula (PSN) at the location of planetary formation, and the delivery mechanism of additional heavy elements to the planets. The locations of the cloud decks also affect the thermal and dynamical structure of Uranus's and Neptune's atmospheres. The abundances of disequilibrium species are expected to change with altitude, and reflect deep atmospheric chemistries as well as the magnitude of convection and vertical mixing. 

This paper describes the main scientific goals to be addressed by the future {\it in situ} exploration of an ice giant. These goals will become the primary objectives listed in a future Uranus or Neptune probe proposal, possibly as a major European contribution to a future NASA ice giant flagship mission. Many of these objectives are within the reach of a shallow probe reaching the 10-bar level. Section \ref{comp} is devoted to a comparison between known elemental and isotopic compositions of Uranus, Neptune, Saturn and Jupiter. We present the different giant planets formation scenarios and the key measurements at Uranus and Neptune that allow disentangling between them. In Section \ref{atmos}, after having reviewed the current knowledge of the atmospheric dynamic and meteorology of the two ice giants, we provide the key observables accessible to an atmospheric probe to address the different scientific issues. Section \ref{mission} is dedicated to a short description of the mission concepts and partnerships that can been envisaged. In Section \ref{pay}, we provide a description of a possible ice-giant probe model payload. Conclusions are given in Section \ref{conc}.

\section{Insights on Uranus and Neptune's Formation from their Elemental and Isotopic Compositions}
\label{comp}

In the following sections, we discuss the constraints that can be supplied by atmospheric probe measurements to the current formation and interior models of Uranus and Neptune. We first discuss the current interior models and the existing elemental and isotopic measurements made in the two giants. We then address the question of the measurement of the key disequilibrium species to assess the oxygen abundance in the two planets, a key element to understand their formation. Finally, we outline the measurement goals and requirements of an atmospheric probe in either of these planets, and how such a mission can improve our understanding of the formation conditions and evolution of these enigmatic worlds.

%__________________________________________________________________
\subsection{Interior Models}
\label{int_mod}
The presence of Uranus and Neptune in our solar system raises the question of how they formed in the framework of the standard theories of planetary formation. Both existing formation models, namely the {\it core accretion} and the {\it disk instability} models, are challenged to explain the physical properties of the two planets. 

In the {\it core accretion} model, the formation of a giant planet starts with the coagulation of planetesimals followed by core growth, concurrent accretion of solids and gas onto the core, and finally by the rapid accretion of a massive gaseous envelope \citep{Mizuno1980,Hubickyj2005,Pollack1996}. If Uranus and Neptune formed at their current orbits, the lower surface density of solids and long orbital periods require that the coagulation of planetesimals proceeds much slower than in the gas giant planet region. Under those circumstances, the ice giants would require formation timescales exceeding the lifetime of the PSN if they accreted {\it in situ} \citep{Bocanegra2015}. In realistic simulations of growth from planetesimals, giant planets cores clear gaps which prevent growth to critical mass before the disk dissipates on $\sim$Myr timescales \citep{Levison2010}. Planetary migration has then been suggested to overcome this issue and might solve the problem \citep{Trilling1998,Alibert2004,Edgar2007,Alexander2009,Helled2014a}. Some help may come from the existence of an outer reservoir of solids in the protosolar disk in the form of pebbles \citep{Lambrechts2012}. \citet{Levison2015} show that this may explain the formation of the giant planets in our Solar System. Note also that Uranus and Neptune probably formed closer to Jupiter and Saturn and then migrated outwards \citep{Tsiganis2005}.

In the {\it disk instability} model, giant planets directly form from gas as a result of gravitational instabilities in a cold disk with a mass comparable to that adopted in the {\it core accretion} model \citep{Boss1997,Mayer2002}. In this case, the growth of disk perturbations leads to the formation of density enhancements in disk regions where self-gravity becomes as important as, or exceeds the stabilizing effects of pressure and shear. To account for their physical properties, it has been proposed that ice giants could consist of remnants of gas giants that formed from disk instability, and whose cores would have formed from the settling of dust grains in the envelopes prior to their photoevaporation by a nearby OB star \citep{Boss2002}.

Furthermore, the interiors of Uranus and Neptune are poorly constrained. A recent study by \citet{Nettelmann2013} based on improved gravity field data derived from long-term observations of the planets' satellite motions suggests however that Uranus and Neptune could present different distributions of heavy elements. These authors estimate that the bulk masses of heavy elements are $\sim$12.5 \Mearth~for Uranus and $\sim$14--14.5 \Mearth~for Neptune. They also find that Uranus would have an outer envelope with a few times the solar metallicity which transitions to a heavily enriched ($\sim$90\% of the mass in heavy elements) inner envelope at 0.9 planet's radius. In the case of Neptune, this transition is found to occur deeper inside at 0.6 planet's radius and accompanied with a more moderate increase in metallicity.

%__________________________________________________________________
\subsection{Uranus and Neptune's Composition}
\label{UN_comp}
The composition of giant planets is diagnostic of their formation and evolution history. Measuring their heavy element, noble gas, and isotope abundances reveals the physico-chemical conditions and processes that led to formation of the planetesimals that eventually fed the forming planets (e.g. \citealt{Owen1999,Gautier2001,Hersant2001}). 

\begin{figure*}[!t]
\begin{center}
\resizebox{\hsize}{!}{\includegraphics[angle=0,scale=1]{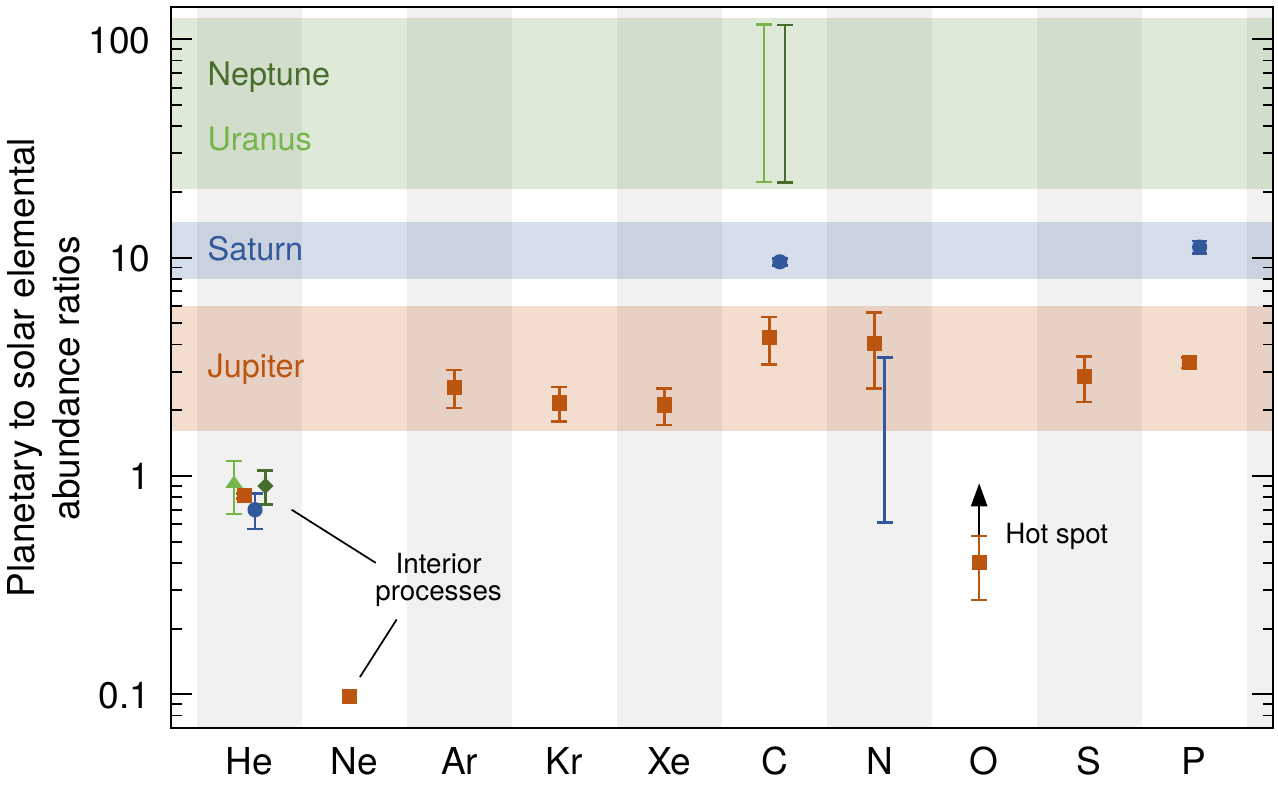}}
\end{center}
\caption{Enrichment factors (with respect to the solar value) of noble gases and heavy elements in the giant planets. See text for references.}
\label{Enrichments} 
\end{figure*}

Heavy element abundances can be derived through a variety of remote techniques (e.g., radio occultation, spectroscopy). However, the most significant step forward regarding our knowledge of giant planet internal composition was achieved with the {\it in situ} descent of the Galileo probe into the atmosphere of Jupiter \citep{Young1998,Folkner1998,Ragent1998,Atkinson1998,Sromovsky1998,Niemann1998,vonZahn1998}. The various experiments enabled the determination of the He/H$_2$ ratio with a relative accuracy of 2\% \citep{vonZahn1998}, of several heavy element abundances and of noble gases abundances \citep{Niemann1998,Atreya1999,Wong2004}. These measurements have paved the way to a better understanding of Jupiter's formation. The uniform enrichment observed in the data (see Figure \ref{Enrichments}) indeed tends to favor a {\it core accretion} scenario for this planet (e.g. \citep{Alibert2005b,Guillot2005}, even if the gravitational capture of planetesimals by the proto-Jupiter formed via {\it disk instability} may also explain the observed enrichments \citep{Helled2006}. On the other hand, the condensation processes that formed the protoplanetary ices remain uncertain, because the Galileo probe probably failed at measuring the deep abundance of oxygen by diving into a dry area of Jupiter \citep{Atreya2003}. Achieving this measurement by means of remote radio observations is one of the key and most challenging goals of the Juno mission \citep{Matousek2007,Helled2014b}, currently in orbit around Jupiter.

At Saturn, the data on composition are scarcer (see Figure \ref{Enrichments}) and have mostly resulted from Voyager~2 measurements and intense observation campaigns with the Cassini orbiter. The Helium abundance is highly uncertain \citep{Conrath1984,Conrath2000,Achterberg2016}, and only the abundances of N, C, and P, have been quantified \citep{Courtin1984,Davis1996,Fletcher2007,Fletcher2009a,Fletcher2009b}. This rarity is the reason why the opportunity of sending an atmospheric probe to Saturn has been studied \citep{Mousis2014a}, and now proposed to ESA and NASA in the M5 and NF4 (respectively) mission frameworks \citep{Mousis2016,Atkinson2016}.

Uranus and Neptune are the most distant planets in our Solar System. Their apparent size in the sky is roughly a factor of 10 smaller than Jupiter and Saturn, which makes observations much more challenging in terms of detectability. This distance factor is probably also the reason why space agencies have not yet sent any new flyby or orbiter mission to either of these planets since Voyager 2. As a consequence, the knowledge of their bulk composition is dramatically low (see Figure \ref{Enrichments}), resulting in a poor understanding of their formation and evolution. To improve this situation significantly enough, we need ground-truth measurements that can only be carried out in these distant planets by an atmospheric probe, similarly to the Galileo probe at Jupiter. In the following paragraphs, we present the current knowledge on the internal composition of the two ice giants (see Tables~\ref{table1} and \ref{table2}), which is mainly inferred from observations of the main reservoirs of the various heavy elements.

\subsubsection{Helium}
The He abundance was first measured by Voyager 2 in both planets during the respective flybys. \citet{Conrath1987,91conrath} report He mass ratios of $Y$$=$0.262$\pm$0.048 and 0.32$\pm$0.05 for Uranus and Neptune, respectively, for an H$_2$-He mixture. \citet{Lodders2009} give a protosolar He mass ratio of 0.278 when considering H$_2$ and He only, leading to the puzzling situation where He was nominally almost protosolar in Uranus and super-protosolar in Neptune. Considering small amounts of N$_2$ in the mixture (with an extreme upper limit of 0.6\% in volume), \citet{Conrath1993} revised the Neptune value down to $Y$$~=~$0.26~$\pm$~0.04, in agreement with the protosolar value. More recently, \citet{Burgdorf2003} have confirmed the value of \citet{Conrath1993}, by constraining the He mass ratio to 0.264$^{+0.026}_{-0.035}$ from far infrared spectroscopy.

All these $Y$ value assume only H$_2$ and He in the gas mixture, as they were derived from measurements all sensitive to atmospheric levels where CH$_4$ was condensed. Below the CH$_4$ cloud base, the CH$_4$ mole fraction is in the range of 1--5\% in both planets (see \ref{sect:carbon}). At those levels, the nominal values of the He mass ratios in Uranus and Neptune then scale to 0.193--0.247 and 0.193--0.247, respectively, when accounting for CH$_4$ (5\% and 1\%, respectively).

In any case, the rather high uncertainty levels on the He abundance makes it difficult to properly constrain interior and evolution models \citep{Guillot2005}, as the error bars still encompass sub- to super-protosolar values. An accurate \textit{in situ} measurement of the He/H$_2$ ratio is thus required to clarify the situation. We note that different datasets and/or different analysis methods never converged to a consensus value for He/H in Jupiter or Saturn from remote sensing only (e.g. \citealt{Conrath1984}, \citealt{Conrath2000}, and \citealt{Achterberg2016} for Saturn). So basically, He/H is achievable from {\it in situ} only.

\subsubsection{Carbon \label{sect:carbon}}
Among heavy element bearing species, only methane has been measured so far in the tropospheres of Uranus and Neptune (and CO in the troposphere of Neptune; \citealt{Marten1993,Moreno2005,05lellouch}). Methane is the main reservoir of carbon at observable levels. However, its deep value remains uncertain because the measurements are inherently more complicated than in the well-mixed atmospheres of Jupiter and Saturn. Methane indeed condenses at the tropopauses of Uranus and Neptune and the observation of its deep abundance cannot be extrapolated from observations probing the stratosphere or the upper troposphere (e.g. \citealt{Lellouch2015}). The first measurements obtained from Voyager-2 radio occultations \citep{Lindal1987,92lindal_nep} and ground-based spectroscopy \citep{95baines} indicate a mole fraction of 2\% in both tropospheres. Coincidentally, these observations all pointed to high latitudes, either because of the ingress/egress latitude of the radio occultation experiments or of the latitudes available from the ground at the time the observations were performed. Interestingly, more recent disk-resolved Hubble Space Telescope observations tend to reveal a more complex situation. \citet{Karkoschka2009,11karkoschka_CH$_4$} and \citet{11sromovsky,14sromovsky} show that the abundance of methane at the equator is twice higher (4$\pm$1\%), and that the high latitude depletion in methane may be caused by meridional circulation and condensation. 

\subsubsection{Nitrogen and sulfur}
N and S are supposedly enriched in the interiors of the ice giants (e.g. \citealt{Owen2003,Hersant2004,Mousis2014b}) and they are carried by ammonia (NH$_3$) and hydrogen sulfide (H$_2$S) in giant planet upper tropospheres. They form a cloud of solid NH$_4$SH already at 30--40 bars, given the low tropospheric temperatures of ice giants. Therefore, the most abundant of the two species will not be entirely consumed by the formation of the NH$_4$SH cloud, and the remaining excess can then be transported up to the condensation level of the corresponding species (between 5 and 10 bars), as illustrated in \citet{Deboer1994}. 

NH$_3$ has been observed in both gas giants and H$_2$S in Jupiter. In Saturn, there are observational hints at the presence of H$_2$S \citep{Briggs1989}. On the other hand, neither of these species has been unambiguously detected in ice giants. Radio-wave observations \citep{depater1989,91depater,Greve1994,Weiland2011} reveal an absorption plateau around 1\,cm wavelength in the brightness temperature spectrum of both planets. NH$_3$ and H$_2$S both have spectral lines in this wavelength range that could result in this broad absorption feature. In Neptune for instance, if it is NH$_3$ that produces the absorption, then its mole fraction is $\sim$10$^{-6}$ between the NH$_4$SH and NH$3$ cloud base levels \citep{91depater}. However, this value is not representative of the deep nitrogen abundance. Similarly, if the centimetric absorption is caused by upper tropospheric H$_2$S, then its mole fraction in the upper troposphere is $\sim$10$^{-4}$ \citep{Deboer1994,Deboer1996}, but is also not representative of the deep sulfur value. To reach such upper tropospheric value, the most recent model requires S to be 10--50 times solar and N $\sim$solar \citep{13Luszczcook}. In both hypotheses, the S/N ratio is found to be super-solar \citep{Deboer1996}. 

Thus, the presumed NH$_4$SH cloud makes measurements of NH$_3$ and/or H$_2$S above the cloud insufficient to constrain the deep N/H or S/H elemental abundances. Uranus and Neptune must be probed at least below the 30 and 50\,bar levels, respectively. However, and following Juno results on NH$_3$ profile retrievals presented in \citet{Bolton2017}, measuring the bulk N and S abundances in Uranus and Neptune may require probing much deeper than the anticipated condensation level of those species.

\subsubsection{Oxygen}
Oxygen is one of the key elements in the formation process of giant planets, as H$_2$O ice was presumably one of the most abundant species in planetesimals beyond the H$_2$O snowline at the time of planet formation. Measuring its precise abundance in the interior of giant planets bears implications on the location where planet formed. The C/O ratio is an important probe in this respect (e.g. \citealt{Ali-Dib2014,Mousis2012,Mousis2014b,Oberg2011,Oberg2016}). The deep O abundance can further help us understand what was the main process that led to the condensation of protoplanetary ices and trapping of other heavy elements. Adsorption on amorphous ice \citep{Bar-Nun1988,Owen1999,Owen2003,Owen2006} and clathration \citep{Lunine1985,Gautier2001,Gautier2005,Alibert2005a,Mousis2006} are the main scenarios described in the literature. They predict large O enrichments, but different in magnitude. The amorphous ice scenario predicts similar enrichments for oxygen and carbon \citep{Owen2003}. On the other hand, the clathration scenario predicts an oxygen abundance $\sim$4 times the carbon abundance \citep{Mousis2014b}.

The temperature profile of Uranus and Neptune has been measured by Voyager 2 radio occultations down to the 2-bar pressure level \citep{Lindal1987,Lindal1990}. Dry or wet adiabatic extrapolation to lower levels shows us that H$_2$O condensation level resides at very high pressure levels of 200--300 bars \citep{13Luszczcook,Cavalie2017}. An atmospheric probe would thus need to reach such depths to measure directly O in Uranus and Neptune. Similar to attempts with Juno at Jupiter, radio waves around 13.5 cm can, in principle, probe down to such depths to characterize the broad absorption from H$_2$O \citep{Matousek2007}. However, the lack of knowledge of the deep thermal lapse rate, especially in the H$_2$O condensation zone, makes it very challenging to disentangle temperature from opacity effects on the radio spectrum of each planet. A third possibility for deriving the deep O abundance consists in measuring the upper tropospheric abundance of a disequilibrium O-bearing species that traces the O abundance at deep levels. Thermochemical modeling then enables deriving the deep O abundance that is responsible for the observed abundance. This indirect approach is presented in more detail in section\,\ref{O_indirect_determination}. So far, it has led to the prediction that the interior of Neptune is extraordinarily enriched in O with respect to the solar value, by a factor of 400 to 600, and that Uranus could be enriched in O by up to a factor of 260 \citep{Lodders1994,13Luszczcook,Cavalie2017}.

\subsubsection{Phosphorus}
Contrary to the gas giant case, ice giant spectra have not yet yielded a detectable levels of PH$_3$ and an upper limit of 0.1 times the solar value was derived by \citet{Moreno2009} in the upper troposphere in the saturation region of PH$_3$. Thus, it is not an upper limit on the deep P/H. The lack of evidence for PH$_3$ in ice giants may be caused by a large deep O/H ratio. \citet{Visscher2005} have shown that PH$_3$ is converted into P$_4$O$_6$ at levels where thermochemical equilibrium prevails. A large O abundance may be the cause of the PH$_3$ depletion in Uranus and Neptune.

%__________________________________________________________________
\subsection{Indirect Determination of Uranus and Neptune's Deep O Abundance}
\label{O_indirect_determination}

Observations of disequilibrium species is one of the methods that can help us complete the determination of the deep elemental composition of giant planets like Uranus and Neptune. Assuming both planets are convective and that their interiors have been fully mixed, we can apply thermochemical modeling in their tropospheres to link upper stratospheric measurements of disequilibrium species to their deep heavy element abundances. The abundances of disequilibrium species are indeed fixed at the level where the timescale of vertical mixing caused by convection becomes shorter than their thermochemical destruction timescale. Using disequilibrium species to estimate the abundance of a deep species is particularly useful in the case of species for which it is very difficult to reach the levels where they are well-mixed. The typical example is O, which is primarily carried by H$_2$O in giant planet deep tropospheres. Observation in the upper troposphere of CO, a disequilibrium species chemically linked to H$_2$O via the net thermochemical reaction CO $+$ 3H$_2$ $=$ H$_2$O $+$ CH$_4$, can thus help us indirectly estimate the deep O abundance by applying thermochemistry and diffusion models.

More or less comprehensive, thermochemical quenching and/or kinetics and diffusion models have been applied to the giant-planet tropospheres in the past decades \citep{Prinn1977,Fegley1985,Fegley1988,Lodders1994,Bezard2002,Visscher2005,13Luszczcook,14cavalie,Wang2016,Cavalie2017}. These models estimate vertical mixing, extrapolate the measured upper tropospheric temperatures to the deep troposphere, and describe the thermochemical reactions at work. Theoretical work describes tropospheric mixing in giant planets \citep{Wang2015} and provides us with estimates. While Neptune with its extraordinarily high tropospheric CO \citep{Marten1993,Marten2005,Guilloteau1993,05lellouch,10lellouch,10fletcher_akari} and very strong internal heat flux \citep{91pearl} is probably fully convective and well-mixed, the very low (or absent) internal heat of Uranus \citep{91pearl} seems to indicate that Uranus is either not fully convective or that it has lost most of its internal heat early in its history (e.g. early giant impact theory, \citealt{Benz1989}). Chemical networks have significantly improved over the last few years \citep{Moses2011,Venot2012}, but there is still space for improvement in the understanding of oxygen chemistry, as shown by \citet{Moses2014} and \citet{Wang2016}. Moreover, the deep tropospheric temperature profile remains quite uncertain. Until very recently, dry or wet adiabatic extrapolations were used (e.g. \citealt{Lodders1994,13Luszczcook,14cavalie}) in giant planet tropospheres. \citet{Guillot1995} and \citet{Leconte2012,Leconte2017} have shown that the situation might be more complex in water-rich interiors, as the temperature profile may significantly depart from adiabatic behavior with the presence of a thin super-adiabatic layer at the H$_2$O condensation level. The influence of such thermal profiles has been explored by \citet{Cavalie2017} in Uranus and Neptune. For a given chemical scheme, they show that applying the new thermal profiles result in much lower O abundances compared to cases where dry/wet adiabats are used. Their nominal models (chemistry, mixing, temperature profile, etc.) show that O is $<$160 times the solar value in Uranus and 540 times solar in Neptune. However, the limitations detailed above remain to be waived for thermochemical and diffusion model results to be more solid. 

CO is not the sole disequilibrium species that can be used to constrain the deep oxygen abundance of giant planets. \citet{Visscher2005} have shown that PH$_3$ is destroyed by H$_2$O in the deep troposphere (in the 1000-bar region ; \citealt{Fegley1985}), following the net thermochemical reaction 4PH$_3$ $+$ 6H$_2$O $=$ P$_4$O$_6$ $+$ 12H$_2$. Measuring the upper tropospheric abundance of PH$_3$ (i.e. below its condensation level) can provide us with a complementary determination of the deep oxygen abundance. To be able to apply this principle to Uranus and Neptune, thermochemical models need to be extended to P species. In this sense, the chemical network proposed by \citet{Twarowski1995} for phosphorus and oxygen species is certainly one starting point, although one would need to validate such a scheme. One would now need to validate such a scheme to the pressure-temperature conditions relevant for Uranus and Neptune deep tropospheres, in the same manner the H-C-O-N  network of \citet{Venot2012} was.

Sending an atmospheric probe to either or both ice giants to measure the upper tropospheric CO and PH$_3$ (below its condensation level) by means of a neutral mass spectrometer, with the aim of constraining the deep O abundance, would undoubtedly boost theoretical and laboratory work to improve current thermochemical models.

%__________________________________________________________________
\subsection{Isotopic Measurements at Uranus and Neptune}
\label{UN_isot}

Table \ref{table3} represents the isotopic ratio measurements realized in the atmospheres of the four giant planets of our solar system. It shows that the only isotopic ratio currently available for Uranus and Neptune is the D/H ratio, which was measured by Herschel-PACS \citep{Feuchtgruber2013}. The case of D/H deserves further {\it in situ} measurements because Herschel observations sampled the pressure in the 0.001--1.5 bar range and deeper sounding could put important constraints on the interiors of Uranus and/or Neptune. The deuterium enrichment as measured by \cite{Feuchtgruber2013} in both planets has been found very close from one another, and its super-solar value suggests that significant mixing occurred between the protosolar H$_2$ and the H$_2$O ice accreted by the planets. Assuming that the D/H ratio in H$_2$O ice accreted by Uranus and Neptune is cometary (1.5--3 $\times $10$^{-4}$), \cite{Feuchtgruber2013} found that 68--86\% of the heavy component consists of rock and 14--32\% is made of ice, values suggesting that both planets are more rocky than icy, assuming that the planets have been fully mixed. Alternatively, based on these observations, \cite{Ali-Dib2014} suggested that, if Uranus and Neptune formed at the carbon monoxide line in the PSN, then the heavy elements accreted by the two planets would mostly consists of a mixture of CO and H$_2$O ices, with CO being by far the dominant species. This scenario assumes that the accreted H$_2$O ice presents a cometary D/H and allows the two planets to remain ice-rich and O-rich while providing D/H ratios consistent with the observations. Deeper sounding with an atmospheric probe should allow investigating the possibility of isotopic fractionation with depth.

The measurement of the D/H ratio in Uranus and/or Neptune should be complemented by a precise determination of $^3$He/$^4$He in their atmospheres to provide further constraints on the protosolar D/H ratio, which remains relatively uncertain. The protosolar D/H ratio is derived from $^3$He/$^4$He measurements in the solar wind corrected for changes that occurred in the solar corona and chromosphere consequently to the Sun's evolution, and to which the primordial $^3$He/$^4$He is subtracted. This latter value is currently derived from the ratio observed in meteorites or in Jupiter's atmosphere. The measurement of $^3$He/$^4$He in Uranus and/or Neptune atmospheres would therefore complement the Jupiter value and the scientific impact of the protosolar D/H derivation. 

The $^{14}$N/$^{15}$N ratio presents large variations in the different planetary bodies in which it has been measured and, consequently, remains difficult to interpret. The analysis of Genesis solar wind samples \citep{Marty2011} suggests a $^{14}$N/$^{15}$N ratio of 441 $\pm$ 5, which agrees with the remote sensing \citep{Fouchet2000} and {\it in situ} \citep{Wong2004} measurements made in Jupiter's atmospheric ammonia, and the lower limit derived from ground-based mid-infrared observations of Saturn's ammonia absorption features \citep{Fletcher2014}. These two measurements suggest that primordial N$_2$ was probably the main reservoir of the atmospheric NH$_3$ present in the atmospheres of Jupiter and Saturn (see \citealt{Owen2001,Mousis2014a,Mousis2014b} for details).  On the other hand, Uranus and Neptune are mostly made of solids (rocks and ices) \citep{Guillot2005} that may share the same composition as comets. N$_2$/CO has been found strongly depleted in comet 67P/Churyumov-Gerasimenko \citep{Rubin2015}, i.e. by a factor of $\sim$25.4 compared to the value derived from protosolar N and C abundances. This confirms the fact that N$_2$ is a minor nitrogen reservoir compared to NH$_3$ and HCN in this body \citep{LeRoy2015}, and probably in other comets \citep{Bockelee2004}. In addition, $^{14}$N/$^{15}$N has been measured to be 127 $\pm$ 32 and 148 $\pm$ 6 in cometary NH$_3$ and HCN respectively \citep{Rousselot2014,Manfroid2009}.  Assuming that Uranus and Neptune have been accreted from the same building blocks as those of comets, then one may expect a $^{14}$N/$^{15}$N ratio in these two planets close to cometary values, and thus quite different from the Jupiter and Saturn values. Measuring $^{14}$N/$^{15}$N in the atmospheres of Uranus and Neptune would provide insights about the origin of primordial nitrogen reservoir in these planets. Moreover, measuring this ratio in different species would enable us to constrain the relative importance of the chemistry induced by galactic cosmic rays and magnetospheric electrons (see \citealt{Dobrijevic2017} for an example in Titan).

The isotopic measurements of carbon, oxygen and noble gas (Ne, Ar, Kr, and Xe) isotopic ratios should be representative of their primordial values. For instance, only little variations are observed for the $^{12}$C/$^{13}$C ratio in the solar system irrespective of the body and molecule in which it has been measured. Table \ref{table3} shows that both ratios measured in the atmospheres of Jupiter and Saturn are consistent with the terrestrial value of 89. A new {\it in situ} measurement of this ratio in Uranus and/or Neptune should be useful to confirm the fact that their carbon isotopic ratio is also telluric. 

The oxygen isotopic ratios also constitute interesting measurements to be made in Uranus and Neptune's atmospheres. The terrestrial $^{16}$O/$^{18}$O and $^{16}$O/$^{17}$O isotopic ratios are 499 and 2632, respectively \citep{Asplund2009}. At the high accuracy levels achievable with meteorite analysis, these ratios present some small variations (expressed in $\delta$ units, which are deviations in part per thousand). Measurements performed in comets \citet{Bockelee2012}, far less accurate, match the terrestrial $^{16}$O/$^{18}$O value. The $^{16}$O/$^{18}$O ratio has been found to be $\sim$380 in Titan's atmosphere from Herschel SPIRE observations but this value may be due to some fractionation process \citep{Courtin2011,Loison2017}. On the other hand, \citet{Serigano2016} found values consistent with the terrestrial ratios in CO with ALMA. The only $^{16}$O/$^{18}$O measurement made so far in a giant planet was obtained from ground-based infrared observations in Jupiter's atmosphere and had a too large uncertainty to be interpreted (1--3 times the terrestrial value; \cite{Noll1995}).

%__________________________________________________________________
\subsection{Volatile Enrichments at Uranus and Neptune
\label{Volatiles}}

\begin{figure*}[!t]
\begin{center}
\resizebox{\hsize}{!}{\includegraphics[angle=0,scale=1]{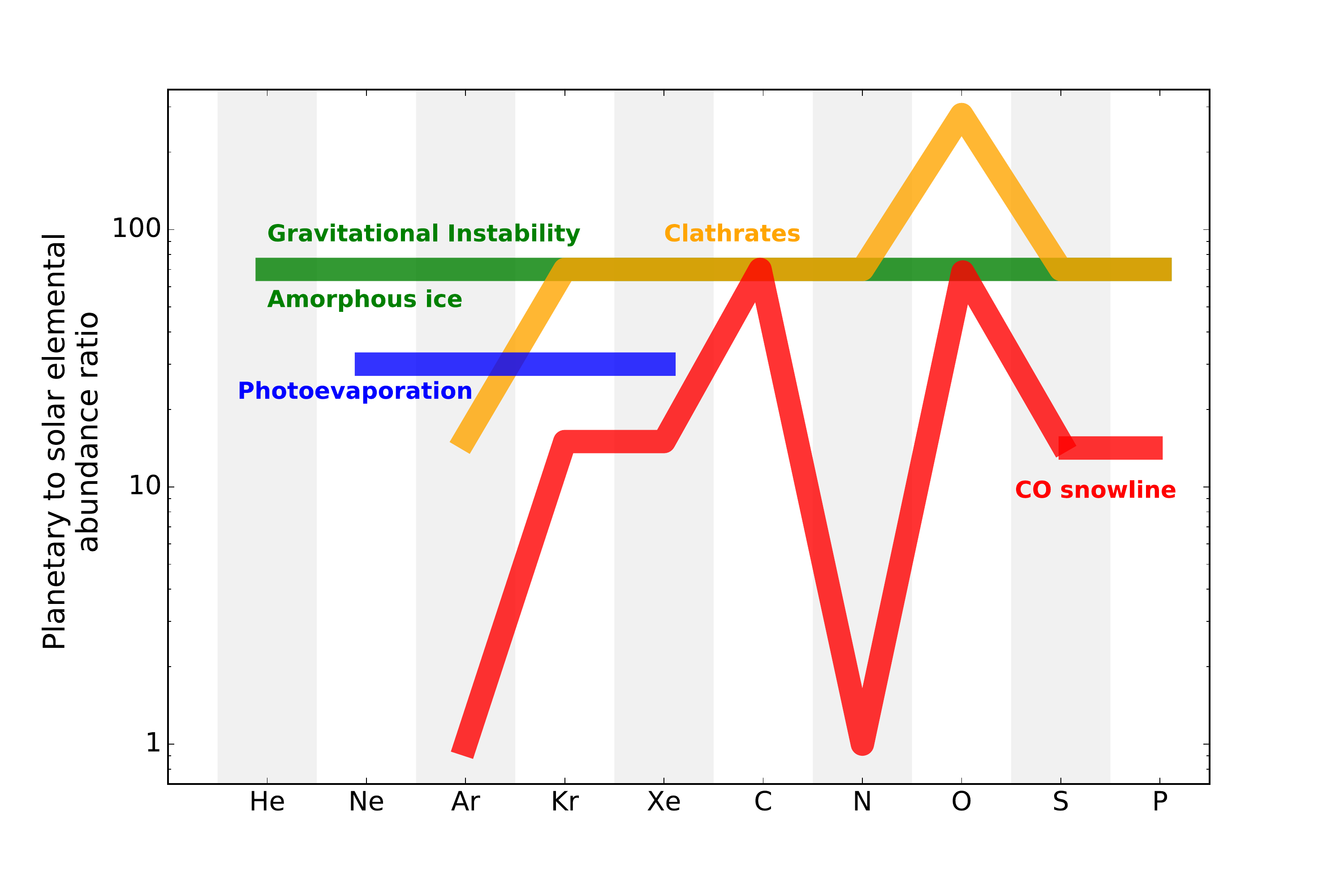}}
\caption{Qualitative differences between the enrichments in volatiles predicted in Uranus and Neptune predicted by the different formation scenarios (calibrations based on the carbon determination). The resulting enrichments for the different volatiles are shown in green (disk instability model and amorphous ice), orange (clathrates), blue (photoevaporation) and red (CO snowline). In their photoevaporation model, \citet{Guillot2006} predict that heavy elements other than noble gases follow the amorphous ice or clathrate predictions. }
\label{enri_pred}
\end{center}
\end{figure*}

The direct or indirect measurements of the volatile abundances in the atmospheres of Uranus and Neptune are key to decipher their formation conditions in the PSN. In what follows, we present the various models and their predictions regarding enrichments in the two ice giants. All predictions are summarized in Figure \ref{enri_pred}.

\subsubsection{Disk Instability Model}
The formation scenario of these planets proposed via the {\it disk instability model}, associated with the photoevaporation of their envelopes by a nearby OB star and settling of dust grains prior to mass loss \citep{Boss2002}, implies that O, C, N, S, Ar, Kr and Xe elements should all be enriched by a similar factor relative to their protosolar abundances in their respective envelopes, assuming that mixing is efficient. Despite the fact that interior models predict that a metallicity gradient may increase the volatile enrichments at growing depth in the planet envelopes \citep{Nettelmann2013}, there is no identified process that may affect their relative abundances in the ice giant envelopes, if the sampling is made at depths below the condensation layers of the concerned volatiles and if thermochemical equilibrium effects are properly taken into account.. The assumption of homogeneous enrichments for O, C, N, S, Ar, Kr and Xe, relative to their protosolar abundances, then remains the natural outcome of the formation scenario proposed by \citet{Boss2002}.

\subsubsection{Core Accretion and Amorphous Ice}
In the case of the {\it core accretion} model, because the trapping efficiencies of C, N, S, Ar, Kr and Xe volatiles are similar at low temperature in amorphous ice \citep{Owen1999,Bar-Nun2007}, the delivery of such solids to the growing Uranus and Neptune is also consistent with the prediction of homogeneous enrichments in volatiles relative to their protosolar abundances in the envelopes, still under the assumption that there is no process leading to some relative fractionation between the different volatiles. 

\subsubsection{Core Accretion and Clathrates}
In the {\it core accretion} model, if the volatiles were incorporated in clathrate structures in the PSN, then their propensities for trapping strongly vary from a species to another. For instance, Xe, CH$_4$ and CO$_2$ are easier clathrate formers than Ar or N$_2$ because their trapping temperatures are higher at PSN conditions, assuming protosolar abundances for all elements \citep{Mousis2010}. This competition for trapping is crucial when the budget of available crystalline water is limited and does not allow the full clathration of the volatiles present in the PSN \citep{Gautier2001,Mousis2012,Mousis2014b}. However, if the O abundance is 2.6 times protosolar or higher at the formation locations of Uranus and Neptune's building blocks and their formation temperature does not exceed $\sim$45K, then the abundance of crystalline water should be high enough to fully trap all the main C, N, S and P--bearing molecules, as well as Ar, Kr and Xe \citep{Mousis2014b}. In this case, all elements should present enrichments comparable to the C measurement, except for O and Ar, based on calculations of planetesimals compositions performed under those conditions \citep{Mousis2014b}. The O enrichment should be at least $\sim$4 times higher than the one measured for C in the envelopes of the ice giants due to its overabundance in the PSN. In contrast, the Ar enrichment is decreased by a factor of $\sim$4.5 compared to C, due to its very poor trapping at 45 K in the PSN (see Figure \ref{enri_pred}). We refer the reader to \citet{Mousis2014b} for further details about the calculations of these relative abundances. 

\subsubsection{Photoevaporation Model}
An alternative scenario is built upon the ideas that (i) Ar, Kr and Xe were homogeneously adsorbed at very low temperatures ($\sim$20--30 K) at the surface of amorphous icy grains settling in the cold outer part of the PSN midplane \citep{Guillot2006} and that (ii) the disk experienced some chemical evolution in the giant planets formation region (loss of H$_2$ and He), due to photoevaporation. In this scenario, these icy grains migrated inwards the disk toward the formation region of the giant planets in which they subsequently released their trapped noble gases, due to increasing temperature. Because of the disk's photoevaporation inducing fractionation between H$_2$, He and the other heavier species, these noble gases would have been supplied in supersolar proportions with the PSN gas to the forming Uranus and Neptune. The other species, whose trapping/condensation temperatures are higher, would have been delivered to the envelopes of Uranus and Neptune in the form of amorphous ice or clathrates. \citet{Guillot2006} predict that, while supersolar, the noble gas enrichments should be more moderate than those resulting from the accretion of solids containing O, C, N, S by the two giants.

\subsubsection{CO Snowline Model}
Another scenario, proposed by \citet{Ali-Dib2014}, suggests that Uranus and Neptune were both formed at the location of the CO snowline in a stationary disk. Due to the diffusive redistribution of vapors (the so-called {\it cold finger effect}; \citealt{Stevenson1988,Cyr1998}), this location of the PSN intrinsically had enough surface density to form both planets from carbon-- and oxygen--rich solids but nitrogen-depleted gas. The analysis has not been extended to the other volatiles but this scenario predicts that species whose snowlines are beyond that of CO remain in the gas phase and are significantly depleted in the envelope compared to carbon. Under those circumstances, one should expect that Ar presents the same depletion pattern as for N in the atmospheres of Uranus and Neptune. In contrast, Kr, Xe, S and P should be found supersolar in the envelopes of the two ice giants, but to a lower extent compared to the C and O abundances, which are similarly very high \citep{Ali-Dib2014}.

%__________________________________________________________________
\subsection{Summary of Key Measurements}
In what follows, we list the key measurements to be performed by an atmospheric entry probe at Uranus and Neptune, in order to better constrain formation and evolution of these planets:

\begin{itemize}[label=\textbullet]

\item Temperature--pressure profile from the stratosphere down to at least 10 bars, because it would help to constrain the opacity properties of clouds laying at or above these levels (CH$_4$ and NH$_3$ or H$_2$S clouds). Around 2\,bars, where CH$_4$ condenses, convection may be inhibited by the mean molecular weight gradient \citep{Guillot1995} and it is thus important to measure the temperature gradient in this region. %Probing deeper than $\sim$40 bars would be needed to assess the bulk abundances of N and S existing in the form of NH$_4$SH but this would require microwave measurements from a Juno-like orbiter, instead of using a shallow probe.

\item Tropospheric abundances of C, N, S, and P, down to the 40-bar level at least (especially for N and S existing in the form of NH$_4$SH clouds), with accuracies of $\pm$10\%~(of the order of the protosolar abundance accuracies). However, these determinations are out of reach of a shallow probe reaching the 10-bar level. Alternatively, N and S could be measured remotely at microwave wavelengths by a Juno-like orbiter. 

\item Tropospheric abundances of noble gases He, Ne, Xe, Kr, Ar, and their isotopes to trace materials in the subreservoirs of the PSN. The accuracy on He should be at least as good as the one obtained by Galileo at Jupiter ($\pm$2\%), and the accuracy on isotopic ratios should be $\pm$1\% to enable direct comparison with other known Solar System values.

\item Isotopic ratios in hydrogen (D/H) and nitrogen ($^{15}$N/$^{14}$N), with accuracies of $\pm$5\%, and in oxygen ($^{17}$O/$^{16}$O and $^{18}$O/$^{16}$O) and carbon ($^{13}$C/$^{12}$C) with accuracies of $\pm$1\%. This will enable us to determine the main reservoirs of these species in the PSN.  

\item Tropospheric abundances of CO and PH$_3$. Having both values puts opposite constraints on the deep H$_2$O \citep{Visscher2005}. CO alone may not be sufficient to enable the evaluation of the deep H$_2$O because of the uncertainties on the deep thermal profile (convection inhibition possible at the H$_2$O condensation level) as shown in \citet{Cavalie2017}.

\end{itemize}

\section{In situ studies of Ice Giant Atmospheric Phenomena}
\label{atmos}

In the following sections, we review the atmospheric dynamics and meteorology of Uranus and Neptune. We explore the scientific potential for a probe investigating atmospheric dynamics and meteorology, clouds and hazes and chemistry. We also provide the key observables accessible to an atmospheric probe to address these different scientific issues.

\subsection{Ice Giant Dynamics and Meteorology}
\label{dynamics}

\subsubsection{Ice Giant Global Winds}

\begin{figure*}
\begin{centering}
\resizebox{\hsize}{!}{\includegraphics[angle=0]{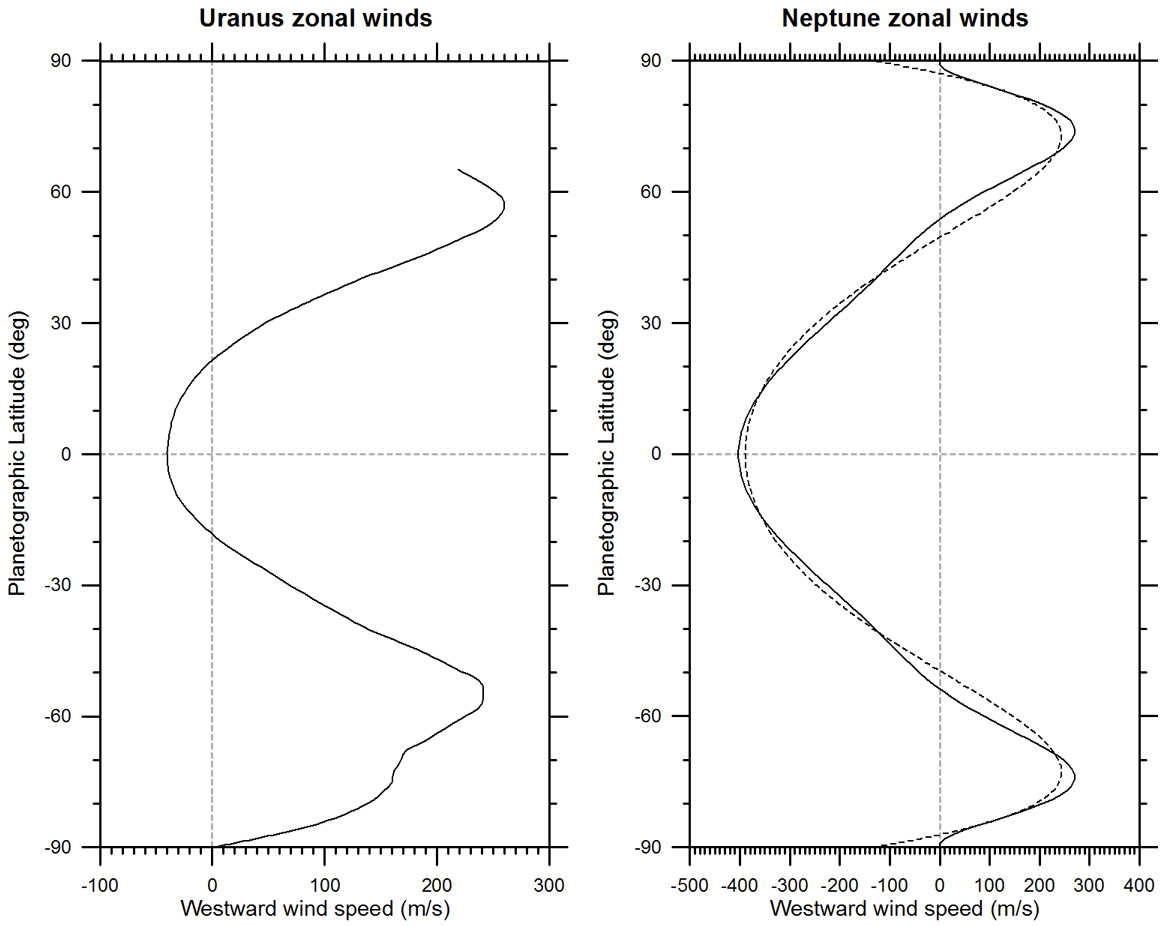}}
\caption{Uranus and Neptune zonal winds. Uranus winds (left panel) combining Keck results from 2012-2014 and a reanalysis of 1986 Voyager images by \citet{15karkoschka} and adopted from \citet{15sromovsky}. Neptune wind (right panel) from Voyager measurements showing different fits to Voyager wind speeds \citep{93sromovsky} and given in \citet{17sanchez}.}
\label{winds}
\end{centering}
\end{figure*}

Uranus and Neptune have zonal winds characterised by a broad retrograde equatorial jet and nearly symmetric prograde jets at high latitudes. Both have very intense winds with Neptune possessing the strongest winds within the Solar System, with its retrograde equatorial jet reaching velocities of -400 m/s and prograde winds at high latitudes reaching velocities of 270 m/s (Figure \ref{winds}). These wind systems are very different to the multi-jet circulations of Jupiter and Saturn with westward equatorial jets.

Winds have been measured on both planets from observations of discrete cloud features gathered by Voyager 2 \citep{86smith,89smith,91limaye,15karkoschka}, Hubble Space Telescope \citep{95sromovsky,01sromovsky,98karkoschka,01hammel} and Keck \citep{05sromovsky,05hammel,09sromovsky,12martin} over multiple decades. The intensity of the winds has appeared to be relatively consistent over time, although there is a large degree of dispersion in the measurements, and it is not clear that the features are genuinely tracking the underlying wind fields \citep[see][for a recent review]{17sanchez}.  

Multi-spectral imaging allows sensing of different cloud altitudes from levels at around 60 mbar to 2 bar \citep{16irwin_ura,16irwin_nep}. Most of the wind analysis show large dispersions with the majority of the observations being sensitive to the upper troposphere (100-200 mbar). It is generally considered that the zonal winds could vary up to 10\% as a consequence of vertical wind shear and tracers at different altitudes. However, the clouds used to track zonal winds may or may not move in the underlying wind fields and large variability is seen \citep{16simon, 16stauffer}.  Long-duration, short-cadence monitoring of light curves of Neptune by Spitzer and Kepler show that the clouds vary on very short time scales.  Similar rapid evolution is seen on the small clouds of Uranus \citep{17irwin}.

{\it In situ} measurements of the deep winds below the observable cloud levels, which are thought to be located at the 2--3 bar level, are key to understanding the nature of the jets on the ice giants. Theoretical models of the origin of atmospheric jets in giant planets are divided in two families: jets could be driven by solar heat flux and shallow atmospheric processes including a crucial role of moist convection in the troposphere \citep[][and references therein]{10lian}; or they could extend deep into the planetary interiors \citep{91suomi,07aurnou}. By monitoring the descent trajectory of an atmospheric probe, in conjunction with measuring the aerosols comprising the visible clouds, we will gain insights into the vertical structure of the ice giant winds for the first time.

\subsubsection{Global Banding, Meridional and Vertical Circulation}
Visible and near-infrared imaging of the ice giants reveal that clouds consist of three types -- zonal banding, discrete bright spots, and dark ovals (see Section \ref{darkovals}).  The zonal bands have low albedo contrast and their meridional extent (5$^\circ$-20$^\circ$ in latitude) is unrelated to the zonal winds and atmospheric temperature structure. In the case of Uranus, since the equinox occurred in December 2007, both hemispheres have been observed at high spatial resolution following the Voyager-2 flyby. The banding distribution was observed in the northern hemisphere in the visible range on Voyager-2 highly processed images \citep{15karkoschka}, and in the southern hemisphere in the red and near-infrared wavelengths \citep{15sromovsky}. Uranus' south polar region extends up to mid latitudes about 45-50$^\circ$S and appears to be bright and featureless. However, the North Pole showed a large number of small-scale bright spots in the near infrared images \citep{15sromovsky}, sugestive of convective motions. The bright spots strongly resemble the cloud pattern seen in the polar regions of Saturn \citep{09delgenio}.

Latitudinally-resolved thermal and compositional data of Uranus and Neptune provide hints of the overall meridional and vertical atmospheric circulation associated with this banded structure. On Neptune, infrared observations from Voyager were interpreted by \citet{91conrath} and \citet{91bezard} in terms of a global circulation system with rising cold air at mid latitudes and overall descent at the Equator and the polar latitudes.  Neptune's summertime pole exhibits a warm vortex in the troposphere and stratosphere that appears bright in the mid-infrared as a consequence of the polar subsidence \citep{07orton_nep,14fletcher_nep}.  The same atmospheric circulation could explain the overall cloud structure in the planet with enhanced storm activity at mid-latitudes, and is consistent with modern infrared and radio-wave observations \citep{14fletcher_nep,13Luszczcook,14depater}.  Uranus exhibits a similar pattern, with cool mid-latitudes and a warm equatorial band in the upper troposphere \citep{87flasar, 15orton}. However, the circulation on both worlds may be much more complex, with suggestions of elevated gaseous abundances at the equator.  The observation that tropospheric methane is enhanced at the equators of both planets compared to the poles \citep{11sromovsky, 11karkoschka_CH$_4$} suggests a different circulation pattern with equatorial upwelling rather than equatorial subsidence.  Ammonia may be similarly enhanced at Uranus' equator \citep{91depater, 03hofstadter}. The nature of ice giant circulation patterns is therefore the subject of considerable debate.  

Intriguingly, the relationship between temperatures, winds and the banded appearance of a giant planet is less clear-cut on Uranus and Neptune than it is on their gas giant cousins.  An atmospheric probe, simultaneously measuring temperatures, winds and aerosol properties, could help to resolve this problem, and to provide insights into the sense of the ice giant circulation patterns. On both Uranus and Neptune, the temperatures in the upper atmosphere are low enough for the equilibration between the ortho- (parallel) and para-hydrogen (anti-parallel) states to play a role in vertical atmospheric dynamics, making measurements of the distribution of the hydrogen ortho-to-para fraction an essential indicator of the global circulation in these planets \citep[e.g.,][]{98conrath}. The ortho-to-para ratio is dependent on temperature and has a long equilibration time. The ortho-to-para ratio affects the overall atmospheric lapse rate and can explain the low heat flux of Uranus \citep{95smith} since Voyager data showed that Uranus' lapse rate and ortho-to-para fraction are not consistent \citep{87gierasch}.  This may indicate thin stratified layers, with fast vertical displacements, such that para-H$_2$ does not get redistributed \citep{85depater, 87gierasch}. In Uranus the ortho to para-H$_2$ ratio varies significantly with both altitude and latitude \citep{98conrath, 03fouchet, 15orton} with a north-south hemispheric asymmetry consistent with the spin-axis tilt of the planet. For Neptune, recent ortho-to-para measurements \citep{14fletcher_nep} suggest that para-H$_2$ disequilibrium is symmetric about the equator, with super-equilibrium conditions at the equator and tropics and at high southern latitudes, and sub-equilibrium conditions at mid-latitudes in both hemispheres. This disequilibrium is consistent with a meridional circulation with cold air rising at mid-latitudes and subsiding at both the poles and the equator, in agreement with other inferences of the global circulation.  

Despite these findings, there exists a degeneracy between measurements of tropospheric temperature, the abundance of helium and the ortho-to-para ratio.  This degeneracy cannot be resolved via remote observations alone, and implies that the vertical para-H$_2$ fraction and its impact on the atmospheric lapse rate is highly uncertain.  An atmospheric probe able to measure each of these parameters simultaneously (as well as determining the helium abundance) would be vital to understand the different sources of energy driving ice giant atmospheric circulations.  Additionally an atmospheric probe would also help resolve uncertainties in remote retrieval of temperatures that assume collision-induced H$_2$ absorption, which depends on the ortho-to-para ratio.

\subsubsection{Meteorology of Uranus and Neptune and Convection}
\label{darkovals}

\begin{figure*}
\begin{centering}
\resizebox{\hsize}{!}{\includegraphics[angle=0]{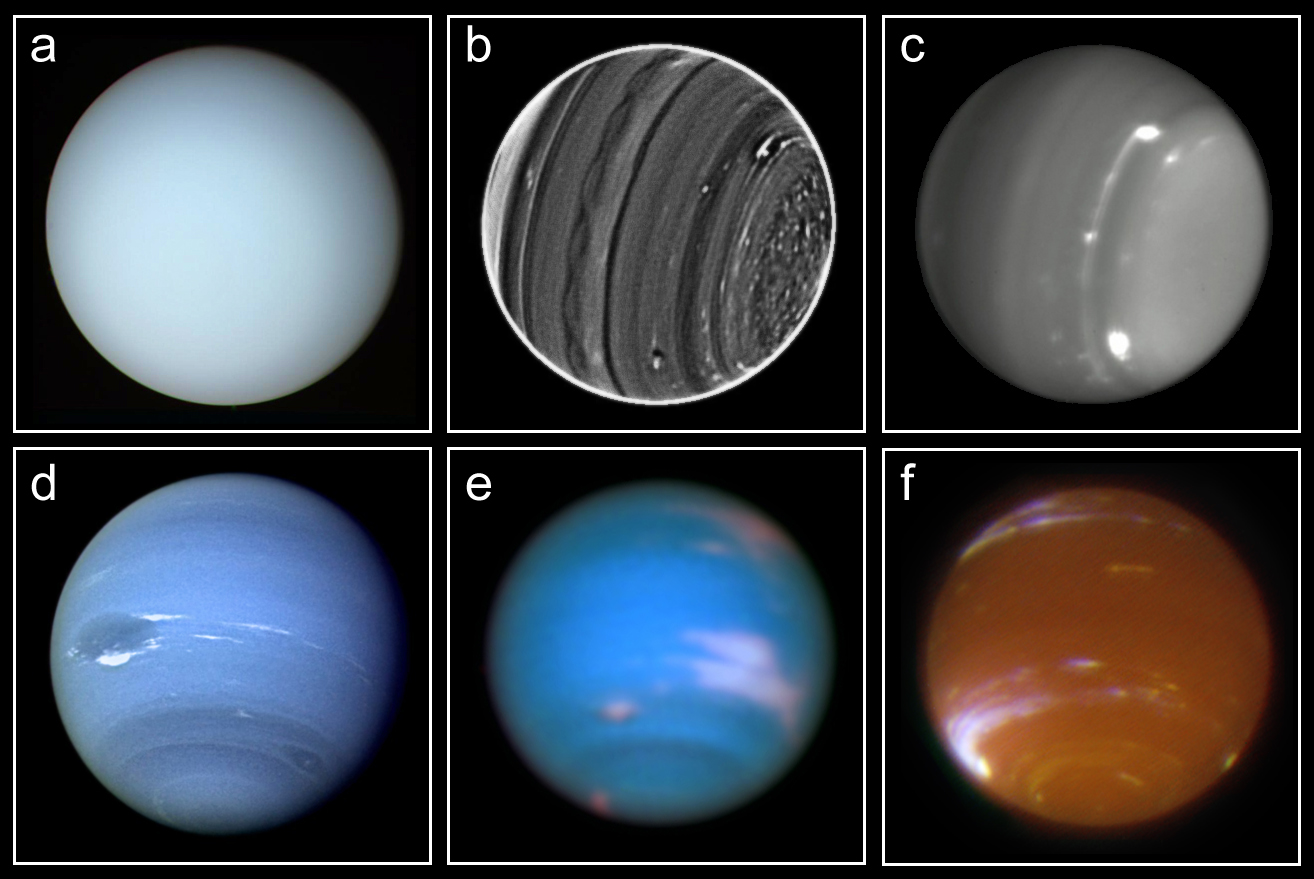}}
\caption{Global views of Uranus and Neptune. Upper row Uranus images in: (a) visible wavelengths from Voyager 2; (b) Near IR with extreme processing of cloud features from \citet{12fry}; (c) Near IR of bright features from \citet{14depater}. Bottom row Neptune images in: (d) visible wavelengths from Voyager 2; (e) Visible wavelengths from HST (image credits: NASA, ESA, and M.H. Wong and J. Tollefson from UC Berkeley); (f) near IR (observations courtesy of I. de Pater).}
\label{images}
\end{centering}
\end{figure*}

The results from an ice giant atmospheric probe would have to be interpreted in light of the different meteorological features that have been observed in Uranus and Neptune.  Figure \ref{images} shows the visual aspect of both planets at a variety of wavelengths from the visible to the near infrared.  Both planets show a recursive but random atmospheric activity at cloud level that can be observed in the methane absorption bands as bright spots \citep{95sromovsky}. Typically, sizes of these features range from 1,000 to 5,000 km. Discrete bright spots are regularly captured at red wavelengths (0.6 - 2.2 $\mu$m) in both planets (but more frequently on Neptune than Uranus). They appear as bright in the methane absorption bands because of their high cloud tops. In Uranus, most of the discrete cloud features are located at the altitude of the methane ice cloud or at deeper levels. The brightest features on Uranus are detected at 2.2 $\mu$m and reach an altitude level of 300--600 mbar, while part of these features are much deeper, being in the lower cloud at 2-3 bars. Uranus's storm activity is more scarce than Neptune's, but can reach a high degree of intensity as occurred in 2014-15 in the latitudes 30$^\circ$-40$^\circ$N \citep{15depater,16irwin_ura,17irwin}. Because of the large obliquity of Uranus, seasonal changes in the cloud and hazes structure are observed, and this requires a long-term survey to determine the altitude where they occur and understand the mechanisms behind their formation under the extremely variable solar insolation conditions.  

Neptune displays both types of discrete cloud activity:  episodic and continuous \citep{94baines, 95sromovsky}. Recently, images taken by the amateur community using improved observing and processing techniques, have been able to capture such features on this planet \citep{17hueso}. On the other hand, the images taken in an ample range of wavelengths from about 400 nm to 2.2 $\mu$m indicate that the clouds are located at higher altitude levels than in Uranus, with cloud tops at around 20-60 mbar whereas other storms are at the $\sim$2 bar level \citep{16irwin_ura, 16irwin_nep}.

This discrete cloud activity could be the result of convective motions, although the sources of energy (ortho-para-H$_2$ conversion, or latent heat release from condensing volatiles) are highly uncertain.  Early models of moist convection on Neptune were examined by \citet{89stoker}, but moist convective storms do not appear to be particularly active on this planet. On Uranus, besides the large long-lived storm system known as the ÒBergÓ \citep{11depater, 15sromovsky}, only a few clouds have been considered as signatures of moist convection in the south polar latitudes \citep{14depater}. However, the relatively low number of high-resolution observations of both planets result in an inability to determine the frequency of moist convective storms in both Uranus and Neptune.

Another way to study moist convective processes is via detections of atmospheric electricity.  Lightning on both Uranus \citep{Zarka1986} and Neptune was detected by Voyager 2, but Neptunian lightning seems weaker, or has a much slower rise time, than Uranian lightning \citep{Gurnett1990,Kaiser1991}. This is unexpected, as Neptune's internal heat source should lead to more convective activity than Uranus. The mechanism for lightning generation is not known, but since both Neptune and Uranus contain clouds of polarizable mixed-phase material such as water and ammonia, then a terrestrial-like mechanism seems possible. Detection of lightning by an atmospheric probe would allow characterisation of the relative strengths and frequencies of lightning, and would enable a deeper understanding of convective and cloud processes at the ice giant planets.

Beyond lightning, atmospheric electrical processes may also contribute to cloud formation at Neptune through ion-induced nucleation producing cloud condensation nuclei, a mechanism first suggested by \citet{92moses}. Ionisation from cosmic rays was closely associated with Neptune's long-term albedo fluctuations by \citet{16aplin}.

Besides the zonal banding and the small-scale bright clouds associated with convective activity, the third most prominent cloud type are larger systems, such as the dark ovals.  Dark oval spots are notable in Neptune where they become conspicuous at blue-green wavelengths. The archetype was the Great Dark Spot (GDS) captured in detail at visible wavelengths in images obtained during the Voyager 2 flyby in 1989 \citep{89smith, 95baines, 98lebeau}. The GDS was first observed at latitude 20$^\circ$S, but after drifting towards the equator it disappeared in about one year. The GDS had a size of 15,500 km (East-West) $\times$ 6,000 km (North-South) and according to the ambient wind profile was an anticyclonic vortex. At least four additional smaller dark vortices have been reported from latitudes 32$^\circ$N to 55$^\circ$S following the Voyager-2 flyby. Bright clouds accompanying the dark ovals are observed at red and near infrared wavelengths and are thought to be the result of air forced upward by the vortex, known as orographic clouds \citep{01stratman}. Other dark spots in Neptune have been observed with similar bright cloud companions, which are thought to develop similarly to orographic clouds by the interaction of the zonal winds with the dark anticyclone.   There is only one report of a dark spot in Uranus similar to Neptune's GDS that was observed in visible wavelengths in 2006 at 28$^\circ$N. It had a size of 1,300 km (North-South) $\times$ 2,700 km (East-West) \citep{09hammel}.

Unlike in Jupiter and Saturn, these large-scale systems can drift meridionally and disappear after a few years moving in the direction of the equator. Some features in Uranus may survive several years like the large ÒBergÓ feature \citep{15sromovsky}. A South Polar Feature in Neptune has been observed since the Voyager observations \citep{11karkoschka} and seems to have a convective origin.  

\subsubsection{Temperature Structure of Uranus and Neptune}
The vertical temperature structure is important as a fundamental constraint on dynamics and chemistry in planetary atmospheres.   Voyager-2 radio-occultation results for Uranus \citep{Lindal1987} and Neptune \citep{92lindal_nep} have provided a sample of the temperature profiles in these atmospheres with a high vertical resolution for a distinct region of each atmosphere. However, as noted above, these results cannot be interpreted in the absence of knowledge of the mean molecular weight, which has been solved simultaneously with simultaneous sensing of infrared radiance in the sampled regions to constrain the bulk composition. This, in turn, relies to some extent on knowledge of the ortho vs. para H$_2$ ratio. Thus it is important to establish all of these for at least one point in the atmosphere to serve as a reference standard for thermal-infrared remote-sensing instruments on a carrier or orbiter, or for more distant remote-sensing observations.  Differences have been noted between the radio occultation results and models for the globally-averaged temperature profile for Uranus \citep[see][and references therein]{14aorton} and Neptune \citep[see][and references therein]{14fletcher_nep}. Thus, remote-sensing observations of the atmospheric probe entry site will be extremely useful to establish the context of the local atmospheric conditions.  This was vital to the interpretation of the Galileo probe entry site, which turned out not to be representative of global particulate and condensate distributions \citep{98orton}. 

To understand energy deposition from external radiance vs internal wave, temperatures in the upper stratosphere through thermosphere are also important. These levels are well above the region to which the radio-occultation measurements are sensitive.  Temperatures are currently characterised only broadly in altitude by a mixture of solar and stellar occultations measured by the Voyager-2 Ultraviolet Spectrometer and ground-based visible observations with large uncertainties and internal inconsistencies \citep{87herbert, 92bishop}.   Measurements by a probe accelerometer will provide substantial information on both upper-atmospheric temperatures, as well as detailed characterisation of gravity waves that contribute to the maintenance of temperatures, as was the case for the Galileo probe \citep{97young}.

\subsubsection{Key Observables of Atmospheric Dynamics}  
Here, we list the key measurements to be made by an atmospheric entry probe at Uranus and Neptune to assess their atmospheric dynamics:

\begin{itemize}[label=\textbullet]
\item Probe descent temperature/density profile. Continuous measurements of atmospheric temperature and pressure throughout the descent would allow the determination of (i) stability regimes as a function of depth though transition zones (e.g., radiative-convective boundary); and (ii) the influence of wave perturbations which could also be used to infer the degree of convection at the probe descent location. The temperature profile is also related to the ortho-to-para ratio and measurements of this ratio as a function of altitude would constrain the degree of vertical convection and the equilibration times of these disequilibrium states.
\item Probe descent accelerometer measurements. Continuous monitoring of the descent deceleration will provide a detailed measurement of the atmospheric density from which the temperature profile can be derived in a region above that of the direct temperature and pressure measurements. 
\item Probe descent winds. Measurements of the vertical profile of the zonal winds from Doppler tracking of an atmospheric probe would provide an insight into the nature of the winds in an ice giant with a small or negligible deep heat source. Doppler wind measurements provide the wind profile in the lower troposphere, well below the tropopause near the region where most of the cloud tracking wind measurements are obtained. Static and dynamic pressures measured from the Atmospheric Structure Instrument (see Section \ref{ASI}) would provide an estimation of the vertical winds, waves, and convection.
\item Conductivity profile.  Measurement of the conductivity profile would indicate what type of clouds support sufficient charge separation to generate lightning. Conductivity measurements combined with meteorological and chemical data (particularly measurements of the physical properties of the aerosols themselves) would also permit extraction of the charge distribution on aerosol particles, and improve understanding of the role of electrical processes in cloud formation, lightning generation, and aerosol microphysics. 
\end{itemize}

Additionally, further measurements during the approach phase would complement the scientific return of the probe:

\begin{itemize}[label=\textbullet]
\item Cloud-tracking observations from a visible to near IR camera or spectral imager on approach could provide a global two-dimensional view of atmospheric dynamics over several weeks at different altitude levels from 2 bar to 60 mbar.  This would allow us to understand the probe descent in the context of nearby meteorological features or changes to the zonal banding.
\item Mid-infrared measurements of the thermal structure and ortho-to-para-H$_2$ ratio distribution at the probe entry site would provide significant information over the global circulation of the planet. 
\item Gravity measurements and deep structure. Measurements obtained by the Voyager 2 flybys imply that the dynamics are confined to a weather layer no deeper than 1,000 km deep in Uranus and Neptune ($\sim$2,000 bar in Uranus and 4,000 bar in Neptune) \citep{13kaspi}. This confinement could be much shallower and information about the deep troposphere below the levels accessible to a probe could be attained by measurements of the gravity field of Uranus and Neptune from the trajectory of a carrier or orbiter.
\item Radio wave detection of lightning from the carrier spacecraft, in addition to optical lighting detections from a camera (dominant emissions are expected to be at 656 nm for Uranus and Neptune), would support the investigation of the conductivity probe.
\end{itemize}

\subsection{Ice-Giant Clouds}
\label{cloud}

\begin{figure*}
\begin{centering}
\resizebox{\hsize}{!}{\includegraphics[angle=0]{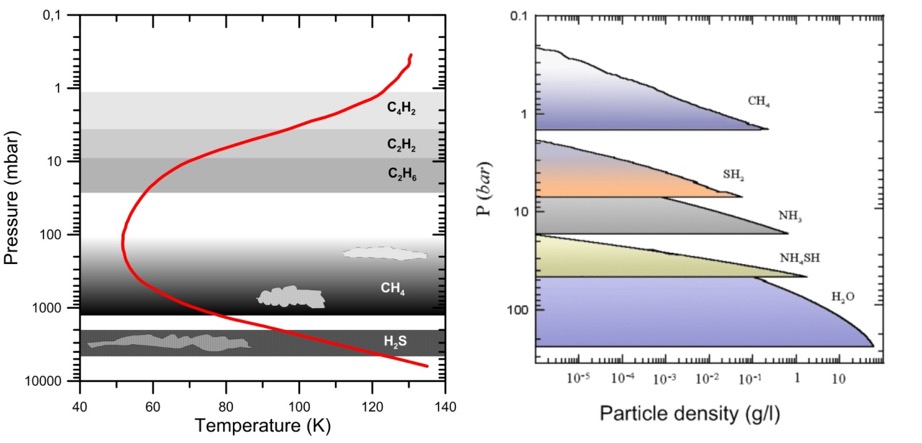}}
\caption{Neptune clouds and hazes. Left: Scheme of the hazes and upper cloud structure accessible to remote sensing, based on those published by \citet{94baines,95baines,09irwin,17irwin}, with temperatures from \citet{92lindal_nep}. Right: Thermochemical model of the main cloud layers in Neptune for the compounds abundances given in the text \citep[following][]{05atreya}. A similar scheme is valid for Uranus.}
\label{clouds}
\end{centering}
\end{figure*}

Our current knowledge of the clouds and hazes on the ice giant planets comes from two main sources: (1) photochemical models of haze and aerosol formation in the upper atmosphere, and thermochemical models based on cloud formation by condensation; (2) analysis of the visible and infrared spectrum by means of radiative-transfer modeling. In the high atmosphere of Uranus and Neptune, methane is photolysed into hydrocarbons (see Section \ref{chemistry}) that diffuse down and condense to form haze layers in the cold stratospheres (altitude range  0.1 to 30 mbar) as the temperature decreases down to $\sim$60 K in the tropopause. The photochemical models suggest the formation of hazes made of H$_2$O, C$_6$H$_6$, C$_4$H$_2$, C$_4$H$_{10}$, CO$_2$, C$_3$H$_8$, C$_2$H$_2$, add C$_2$H$_6$ from top to bottom \citep{88romani, 93romani, 91west, 94baines, 95baines, 95moses, 05moses, 10dobrijevic, 17moses}, where the oxygen species derive from external sources such as interplanetary dust or comets (Figure \ref{clouds}). 

Thermochemical equilibrium cloud condensation (ECC) models are based on the vertical temperature and composition distributions. They give the altitude of the formation of the cloud bases and the vertical distribution of the density in the cloud according to the different species that condense and following the saturation vapor pressure curves based on the Clausius-Clapeyron equation \citep{Sanchez2004, 05atreya} (Figure \ref{clouds}). Depending on the abundances of the condensables, at least five cloud layers are predicted to form. For deep abundances relative to the solar value of O/H = 100, N/H = 1, S/H = 10 and C/H = 30--40, four cloud layers of ice particles of CH$_4$, H$_2$S, NH$_4$SH, H$_2$O form between pressure levels  0.1 bar and 50 bar (representing a vertical distance of about 500 km, Figure \ref{clouds}). The lower water-ice cloud is at the top of a massive aqueous water cloud that could extend down to 1,000 bars or more. It should be noted, however, that the existence of a H$_2$S cloud depends upon sulphur being more abundant than nitrogen on the ice giants.  Although this depletion of nitrogen has been suggested by microwave observations, it remains extremely uncertain, and there is a possibility that an NH$_3$ ice cloud could form if N is more abundant than S, as on Jupiter and Saturn.  An atmospheric probe penetrating down to 50--100 bar should sense and measure the properties of all these cloud layers, whereas a shallow probe to 10 bar would reach the H$_2$S cloud.

Visible and near-infrared images of Uranus and Neptune, combined with their reflectance spectra analysed via radiative-transfer models show that, to first order, the structure and properties of the accessible clouds in both Uranus and Neptune are similar. They consist of an extended haze with top at 50-100 mbar located above a thin methane cloud of ice condensates with its base at 1.3 bar. This cloud is above another cloud of H$_2$S ice that is thin in thickness but optically thick that is located between 2 and 4 bar or pressure, presumed to be formed by H$_2$S condensates \citep[][and references therein]{89hammel,09irwin}. This model, consisting of two cloud layers and an extended haze, has been proposed based on many independent studies, the more recent ones by \citet{13tice,15dekleer,16irwin_ura,16irwin_nep}. The effective radius for the stratospheric haze particles is  0.1-0.2 $\mu$m and of 1-1.5 $\mu$m for the methane tropospheric cloud \citep{91west,94baines,17irwin}.  It should be noted, however, that these inferences from radiative transfer modelling are degenerate, with multiple possible solutions for the optical properties (e.g., aerosol composition and refractive indices) and vertical structure.  Furthermore, they are being updated all the time as new sources of laboratory data for the cloud and methane absorptions become available.  An atmospheric probe would directly test the results of these remote observations, measuring the properties of the aerosols as a function of depth to provide a ground-truth to remote sensing observations, and accessing clouds much deeper than possible from remote platforms.    

\subsubsection{Key Observables of Ice Giant Clouds}
The clouds of an ice giant are the filter through which remote observations attempt to determine their bulk composition.  An atmospheric probe would allow us to constrain the vertical structure and physical properties of the aerosols responsible for the planet's appearance in reflected sunlight, as well as revealing the relationship between the atmospheric lapse rate, gaseous composition, and the resulting aerosols.  Key measurements from the atmospheric probe include:

\begin{itemize}[label=\textbullet]
\item Determinations of the properties of the clouds and hazes along the descent path, measuring the scattering properties at a range of phase angles, the number density as a function of depth, the aerosol shape and opacity properties.    Each of these measurements would help constrain the aerosol composition.
\item Determine the influence of cloud condensation or photochemical haze formation on the temperature lapse rate, and deduce the amount of energy relinquished by this phase change.
\item Determine the effect of cloud formation on the vertical profiles of key condensable species (CH$_4$, NH$_3$, H$_2$S).
\end{itemize}

\subsection{Ice-Giant Chemistry}
 \label{chemistry}
Section \ref{comp} provided an overview of the bulk chemical composition and thermochemistry of Uranus and Neptune, revealing that of the primary elements heavier than hydrogen and helium (namely carbon, nitrogen, oxygen, sulphur and phosphorus), only carbon has been definitively detected in remote sensing observations in the form of methane and CO.  The key cloud-forming volatiles -- NH$_3$, H$_2$S and H$_2$O -- remain largely inaccessible to remote sensing, and we have only upper limits on disequilibrium species such as PH$_3$.  The chemistry of the upper tropospheres and stratospheres of the ice giants is a product of the source material available, as we describe in the following sections.  An atmospheric probe must be able to measure the vertical distributions of gaseous species and aerosols to determine the chemical processes at work on the ice giants, allowing us to contrast (i) the implications of different photochemical mixing efficiencies between Uranus and Neptune; and (ii) the different physical and chemical processes at work on the gas and ice giants.  Compositional differences between these hydrogen-dominated atmospheres can result from many factors, including \citep{05moses}:  differences in photolytic rates due to different heliocentric distances; different reaction rates and condensation due to different atmospheric temperatures; different strengths of atmospheric mixing; differences in auroral energy and potential ion-neutral chemistry; and different influxes of material of exogenic origins.  Understanding the importance of these different influences requires a robust, direct measurement of ice giant chemistry.
 
\subsubsection{Methane Photochemistry}
Despite containing significantly more tropospheric methane than the gas giants \citep[up to $\sim$4\% in mole fraction at low latitudes,][]{14sromovsky, 11karkoschka_CH$_4$}, the cold temperatures of the ice giant tropopause forces methane to condense, acting as an effective cold-trap.  However, some methane gas is able to escape into the stratosphere, either via convective overshooting or slow diffusion through warmer regions \citep[e.g.,][]{07orton_nep}, where it helps to heat the stratosphere via solar absorption in the near-infrared, yielding the stratospheric inversions on Uranus and Neptune.  Once in the stratosphere, ultraviolet photolysis of methane initiates a chain of photochemical reactions to generate heavier hydrocarbons \citep{83atreya,89summers,89romani_nep,92bishop,05moses,10dobrijevic} which dominate the mid-infrared emission spectra observed from Earth-based and space-based facilities \citep[e.g., ISO, AKARI and Spitzer;][]{97feuchtgruber,98encrenaz,06burgdorf,08meadows,10fletcher_akari,14borton}, and produce absorptions in UV occultation observations from Voyager \citep[e.g.,][]{87herbert,90bishop}.     

\begin{figure*}
\begin{centering}
\resizebox{\hsize}{!}{\includegraphics[angle=0]{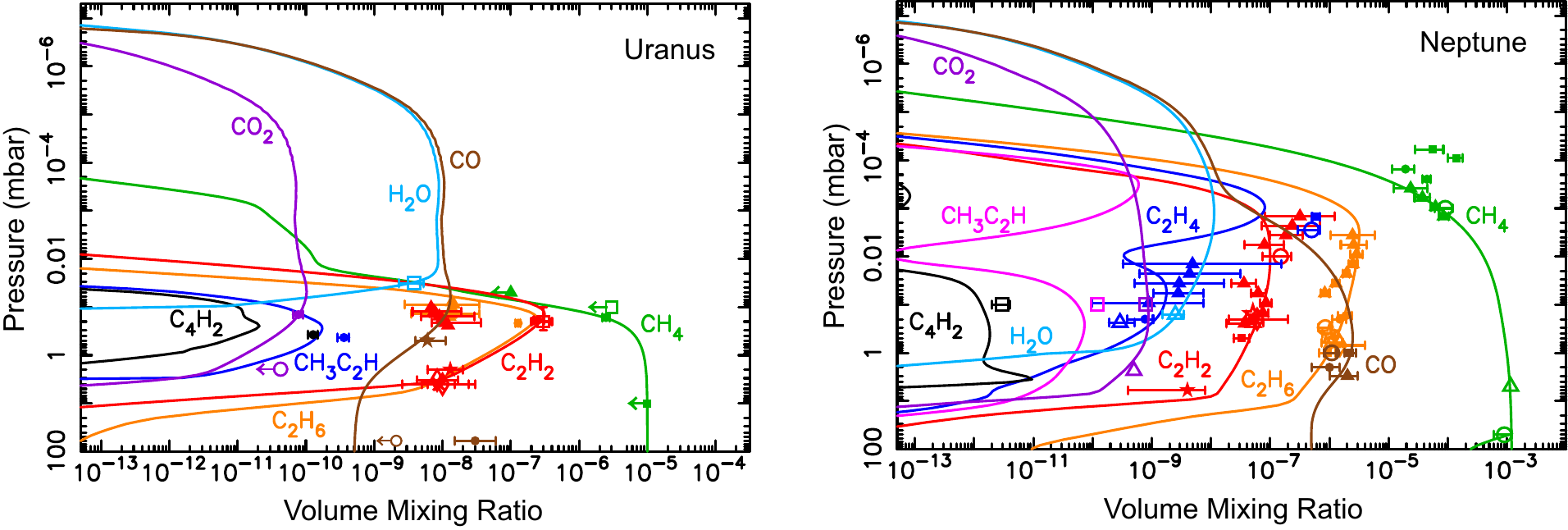}}
\caption{Comparison of the vertical distributions of hydrocarbons and oxygen compounds in the stratospheres of Uranus (left) and Neptune (right), following \citet{17moses}.  Points with error bars are measurements from a wide variety of literature sources - see \citet{17moses} for full details.  The difference in homopause altitudes, driven by the different efficiencies of vertical mixing, cause significant differences in the stratospheric chemistry.}
\label{moses_chem}
\end{centering}
\end{figure*}

Species detected on both planets so far (Figure \ref{moses_chem}) include ethane (C$_2$H$_6$), acetylene (C$_2$H$_2$), methylacetylene (C$_3$H$_4$) and diacetylene (C$_4$H$_2$) \citep[e.g.,][]{06burgdorf, 14borton,08meadows,10fletcher_akari}, whereas ethylene (C$_2$H$_4$) and methyl (CH$_3$) have only been detected on Neptune.  Some species, such as propane (C$_3$H$_8$) and benzene (C$_6$H$_6$) remain undetected due to the difficulties of separating their emissions from bright nearby features. The brightness of a particular emission feature is determined by both the stratospheric temperature profile and the vertical gaseous distribution, the latter of which is shaped by the strength of vertical mixing (e.g., upward diffusion and slow settling), the net chemical production rate profile, the altitude of the photolysis region, and the possibility of condensation of the hydrocarbons to form haze layers.  Measuring temperature and composition remotely is a degenerate problem, and for the species listed above we rarely have any confidence in the measured vertical profiles.  Furthermore, these profiles are likely to vary with latitude if methane is more elevated at the equator due to enhanced vertical mixing, or at the poles if CH$_4$ leaks through warm polar vortices \citep{89yelle,11greathouse,14fletcher_nep}, and some species are observed to vary with time \citep[e.g., Neptunian ethane,][]{06hammel,14fletcher_nep}.  Indeed, hydrocarbon production rates depend on solar insolation and will be seasonally variable, with maximum abundances expected in the summer hemisphere in the absence of circulation.  

Atmospheric circulation, either via large-scale inter-hemispheric transport as part of some global circulation pattern, or via general diffusive mixing, is expected to generate observable differences in the methane photochemistry between Uranus and Neptune (Figure \ref{moses_chem}).  Uranian mixing appears more sluggish, meaning that CH$_4$ will not reach such high stratospheric altitudes as on Neptune \citep[i.e., a low methane homopause,][]{87herbert,90bishop}, therefore ensuring that photochemistry on Uranus occurs in a different physical regime (higher pressures) than on any other giant planet, suppressing photochemical networks \citep{91atreya}. This difference can be readily seen in the ratio of ethane to acetylene, which is much larger than unity on Jupiter, Saturn and Neptune, but smaller than unity on Uranus \citep{05moses, 14borton}. \citet{14borton} use Spitzer mid-infrared observations of Uranus to demonstrate that the slow vertical mixing implies that the hydrocarbons are confined to altitudes below the 0.1-mbar pressure level.  Furthermore, they suggest that there is no evidence for an increase in mixing (and therefore hydrocarbon abundances) near Uranus' 2007 equinox, despite suggestions of an increase in dynamical activity in the troposphere at this time (see Section 3.1).  An atmospheric probe, able to distinguish the vertical profiles of stratospheric temperature and hydrocarbon composition (and to potentially detect previously-undetected species), would allow the first robust tests of stratospheric chemistry models \citep[e.g.,][]{05moses,14borton} balancing the competing influences of seasonal photochemistry, vertical mixing and aerosol condensation at work within an ice giant stratosphere.

\subsubsection{Exogenic Species}
Section \ref{O_indirect_determination} described the potential internal source of CO as a disequilibrium species on Uranus and Neptune and bulk H$_2$O as a volatile species hidden deep below the reaches of remote sensing. But H$_2$O, CO and CO$_2$ are also present in ice giant stratospheres from external sources (Figure \ref{moses_chem}), such as cometary impacts, satellite debris or ablation of interplanetary dust grains and micrometeoroids \citep[e.g.,][]{97feuchtgruber, 05lellouch, 16poppe, 17moses}.  Stratospheric water was detected by ISO \citep{97feuchtgruber}; CO from the fluorescent emission in the infrared \citep{04encrenaz, 10fletcher_akari} and sub-millimeter emission \citep{93martin,05lellouch,07hesman,10lellouch,14cavalie}; Uranus' CO$_2$ from Spitzer \citep{06burgdorf,14borton} and Neptune's CO$_2$ from ISO \citep{97feuchtgruber}.  These oxygenated species can therefore play a part in the photochemical reaction pathways along with the methane photolysis described above.  The relative abundances of these three species can provide clues to their origins \citep{14cavalie,14borton,17moses}. 

The vertical distribution of H$_2$O and CO$_2$ is not expected to differ significantly between the two planets.Ê However, the oxygen-related chemistry on Uranus is anomalous because the methane homopause is so low that there is not a very large interaction region between the hydrocarbons and oxygen species before the H$_2$O condenses, in comparison to Neptune, so there should be less coupled oxygen-hydrocarbon photochemistry \citep[e.g.,][]{17moses}. Neptune is anomalous because CO is significantly enriched in the upper stratosphere, which likely comes from a large cometary impact \citep{05lellouch, 07hesman,13Luszczcookb,17moses}.  Oxygenated species play other roles in shaping the stratospheric structure:  CO and CO$_2$ would be photolysed and play a role in the photochemistry at high altitude, potentially leading to a secondary peak of hydrocarbon production above the methane homopause level, and therefore influencing the thermal structure (via excess heating/cooling).  Water may condense to form high-altitude haze layers.  Finally, stratospheric HCN and CS can become involved in the chemistry of the stratosphere, potentially originating from large cometary impacts \citep{Lellouch2015}. HCN can also originate from galactic-cosmic-ray-induced chemistry of intrinsic N$_2$ from the interior, or photochemistry of nitrogen flowing in from Triton \citep[e.g.,][]{94lellouch}.  A direct measurement of the vertical distribution of these upper stratospheric compounds would shed light on their origins and importance in shaping the conditions in the upper stratospheres of the ice giants.  

\subsubsection{Tropospheric Photochemistry}
Disequilibrium species are those that are detectable in a giant-planet upper troposphere as a result of vigorous vertical mixing.  At some pressure deep in the troposphere (the Òquench levelÓ), the rate of vertical mixing becomes faster than the rate of thermochemical destruction and the abundance becomes frozen in at a value representing the quenched equilibrium composition \citep{Fegley1985}.  On the gas giants Jupiter and Saturn, this provides detectable amounts of phosphine (PH$_3$), CO, arsine (AsH$_3$) and germane (GeH$_4$) in their upper tropospheres \citep[e.g.,][]{04taylor, 15fletcher}.  As described in Section \ref{O_indirect_determination}, only CO has been observed on the ice giants, with no detections of the other potential disequilibrium species.

However, on Jupiter and Saturn the primary condensable (NH$_3$) and disequilibrium molecule (PH$_3$) have vertical profiles that are significantly altered by the coupled tropospheric photochemistry \citep[e.g.,][]{84atreya}.  The same could also be true of H$_2$S, AsH$_3$ and GeH$_4$ \citep{Fegley1985}.  Unfortunately, little is known about the reaction pathways for these tropospheric constituents, but the works of \citet{84kaye} and \citet{09visscher} suggest that a variety of photo-produced species could exist, including diphosphine (P$_2$H$_4$), hydrazine (N$_2$H$_4$), and gas-phase N$_2$.  Diphosphine and hydrazine may condense to form a part of the hazes observed on Jupiter and Saturn, and photo-processing of these species may contribute to the arrays of observable colours.  These hazes have a feedback effect on the chemistry, sometimes shielding the UV photolysis of deeper gas molecules, and implying that the vertical distribution of gases above the clouds are sensitive to the strength of transport, condensation, and the efficiency of the photochemistry.   If these species (primarily NH$_3$, H$_2$S and PH$_3$) can be definitively identified by an atmospheric probe, then their vertical profiles would reveal much about the competing transport and chemistry processes at work. This is essential before their deep abundances can be used to constrain the bulk composition of these planets in Section \ref{O_indirect_determination}.

\subsubsection{Key Observables for Atmospheric Chemistry}
Section \ref{chemistry} has described the rich array of molecular species and aerosols that could be present on the ice giants as a result of photochemistry of the source material.  The vertical distribution of the source materials (methane, oxygen and nitrogen compounds, or disequilibrium species) depend on the nature of their delivery, from vertical mixing, large-scale circulation or external influx.  Some of these source materials and their products are challenging to observe remotely.  Even if their spectral features are identifiable, there remains a fundamental degeneracy between the vertical temperature and composition that prevents a comprehensive understanding of the processes involved.  Key measurements providing a ground-truth for these remote sensing measurements include:

\begin{itemize}[label=\textbullet]
\item Vertical profiles of atmospheric temperature and lapse rates from the stratosphere into the troposphere.
\item Multiple direct measurements of atmospheric composition as a function of altitude to determine photochemical source regions, homopause altitudes, condensed phases and the influence of the cold trap.
\item First detections of precursor molecules (e.g., PH$_3$, NH$_3$, H$_2$S), their photochemical products, and constraints on their vertical profiles.
\item Vertical distribution of aerosols produced via condensation of photolytic products.
\end{itemize}
 
A key challenge for an atmospheric probe to study atmospheric chemistry is the need to track the thermal structure and chemical composition from high altitudes, down through the tropopause and into the cloud-forming region.  

\subsection{Atmospheric Phenomena Summary}
A single entry probe descending into the atmosphere of an ice giant would provide significant new insights into the physical and chemical forces shaping their observable atmospheres.  In addition to providing ground-truth for the parameters that can be crudely measured remotely -- the thermal structure, the gaseous abundances above the clouds, the windspeeds at the cloud-top, and the vertical aerosol structures -- the probe would provide a wealth of insights into properties that are inaccessible.  These include measuring gaseous species that are hidden deep below the cloud layers; determining the roles of cloud condensation, vertical mixing and photochemistry in shaping the vertical distributions of trace species; and measuring temperatures and winds deep below the clouds.  The ice-giant probe measurements will allow the first direct and unambiguous comparison with the Galileo probe results at Jupiter, to see how the thermal structure, composition, clouds and chemistry differ between the gas and ice giants of our solar system.

\section{Proposed mission Configuration and Profile}
\label{mission}

\subsection{Probe Mission Concept}
\subsubsection{Science Mission Profile}

To measure the atmospheric composition, thermal and energy structure, clouds and dynamics requires {\it in situ} measurements by a probe carrying a mass spectrometer (atmospheric and cloud compositions), atmospheric structure instrument (thermal structure and atmospheric stability), nephelometer (cloud locations and aerosol properties), net flux radiometer (energy structure), and Doppler-wind experiment (dynamics). The atmospheric probe descent targets the 10-bar level located about 5 scale heights beneath the tropopause. The speed of probe descent will be affected by requirements imposed by the needed sampling periods of the instruments, particularly the mass spectrometer, as well as the effect speed has on the measurements. This is potentially an issue for composition instruments, and will affect the altitude resolution of the Doppler wind measurement. Although it is expected that the probe batteries, structure, thermal control, and telecom will allow operations to levels well below 10 bars, a delicate balance must be found between the total science data volume requirements to achieve the high-priority mission goals, the capability of the telecom system to transmit the entire science, engineering, and housekeeping data set (including entry accelerometry and pre-entry/entry calibration, which must be transmitted interleaved with descent data) within the descent telecom/operational time window, and the probe descent architecture which allows the probe to reach 10 bars.

\subsubsection{Probe Mission Profile to Achieve Science Goals}
A probe to Uranus or Neptune will be carried as one element of a dedicated ice-giant exploration, likely a NASA flagship mission \citep{Elliott2017}. The probe is designed for atmospheric descent under parachute to make measurements of composition, structure, and dynamics, with data returned to Earth using the Carrier Relay Spacecraft (CRSC) as a relay station that will receive, store, and re-transmit the probe science and engineering data. While recording the probe descent science and engineering data, the CRSC will make radio-science measurements of both the probe relay link signal strength from which abundances of key microwave absorbers in Uranus's atmosphere can be retrieved, and probe relay link frequency from which Doppler tracking of the probe can be performed to retrieve the atmospheric dynamics. 

Upon arrival in the vicinity of the ice giant system, the atmospheric probe will be configured for release, an extended coast, entry, and the atmospheric descent mission. For proper probe delivery to the entry interface point, the CRSC with probe attached is placed on a planetary-entry trajectory, and is reoriented for probe release. The probe coast timer and pre-programmed probe descent science sequence are loaded prior to release from the CRSC, and following a spin-up period, the probe is released for a ballistic coast to the entry point. It is beneficial to Doppler track the CRSC prior to, during, and subsequent to the release event, so that the observed change in CRSC speed can help reconstruct the probe release dynamics and reduce the uncertainty in the probe arrival location. If feasible, it is also beneficial to image the probe from the CRSC shortly after probe release. Optical navigation of the probe relative to background stars can help reduce the uncertainty in the probe release dynamics, departure trajectory, and arrival location. Following probe release, a deflect maneuver is performed to place the CRSC on the proper overflight trajectory for the probe descent relay communications. An important consideration during probe coast is to ensure that probe internal temperatures remain within survival range by careful thermal design and management, and, as needed, by batteries. It is important to recognize an important trade exists between a probe release closer to the planet (deeper within the planet's gravity well) resulting in a shorter coast period with less impact on probe thermal control requirements, power, and required battery complement, as well as a smaller uncertainty in probe entry interface location but at a cost of a higher $\Delta$$V$ (and therefore more fuel) for the CRSC, vs. an earlier release requiring a smaller CRSC deflection $\Delta$$V$ and less fuel, but requiring a longer coast, a larger uncertainty in probe-interface arrival location, and a more significant impact on probe thermal and power. During the coast period the probe will periodically transmit beacons to the CRSC to provide probe coast survival and overall health status. However, once released from the CRSC there is no opportunity to send commands to the probe. 

Prior to arrival, the probe coast timer awakens the probe for sequential power-on, warm-up, and health checks of subsystems and instruments, and to perform preliminary instrument calibrations. One of the first systems to be powered on is the ultrastable oscillator that requires an extended warmup period to achieve operational stability needed to support the Doppler Wind Experiment. Although all instruments are powered on for warmup and calibration, the only instrumentation collecting data during entry will be the accelerometers located at the probe center of mass to measure the entry accelerations required to reconstruct the probe entry trajectory and to retrieve the density profile of the upper atmosphere. The accelerometers provide a g-switch trigger to initiate parachute deployment and configure the atmospheric probe for its descent science mission. The parachute sequence is initiated above the tropopause by firing a mortar through a breakout panel in the aft cover and deploying a pilot parachute. The pilot parachute pulls off the probe aft cover while extracting the main descent parachute. After a short period of time, the probe heatshield will be released and the probe will establish a communication link with the CRSC and commence descent operations. The need for probe rotation during descent is not yet well defined, but spin vanes to control minimum and maximum spin rates and sense will be carefully studied.

Under the parachute, any required mode changes in descent science operations with altitude can be guided by data from the Atmospheric Structure Instrument pressure and temperature sensors, thereby providing the opportunity to optimize the data collection for changing science objectives at different atmospheric depths. To satisfy mission success criteria the probe science data collection and relay transmission strategy will be designed to ensure the entire probe science data set is successfully transmitted to the CSRC before the descent probe reaches the targeted depth. Data collected beyond the target depth will be returned as long as the relay link survives.

The actual descent sequence and timing, main parachute size and descent speeds, and time to reach the required depth (nominally 10 bars) will depend upon considerations of instrument science data generation and total data volume to be returned. During descent, the probe science payload will make measurements in real time, with data buffered for later return. The probe pre-entry and entry instrument calibration, probe housekeeping, and entry accelerometry data must also be returned, and is interleaved with the probe descent science and required engineering/housekeeping data. The probe telecom system will comprise two cross-polarized channels separated slightly in frequency, with each channel nominally transmitting identical data sets for redundancy. If extra bandwidth is required, it is possible to transmit high-priority science and engineering data on both channels, and to separate lower priority data between the two channels. To reduce the possibility of data loss during brief relay link dropouts, the option exists to provide a slight time offset of the two channels. The probe descent mission will likely end when the telecom geometry becomes so poor that the link can no longer be maintained, when the probe reaches a depth that the overlying atmospheric opacity is so large that the link cannot be supported, or when battery depletion or increasing thermal and/or pressure effects cause systems in the vented probe to fail.

The CRSC receives the probe data, storing multiple copies in redundant on-board memory. At the completion of the probe descent mission, the CRSC reorients to point the High-Gain Antenna towards Earth and the multiple copies of the probe science and engineering data are downlinked. 

\subsection{Probe Delivery}
\subsubsection{	Interplanetary Trajectory}
Four characteristics of interplanetary transfers from Earth to Uranus or Neptune are of primary importance: the launch energy, the duration of the transfer, the $V_\infty$ of approach (VAP) to the destination planet, and the declination of the approach asymptote (DAP). The higher the launch energy, the smaller the mass a given launch vehicle can deliver to that energy. The duration of the transfer is of particular interest for Uranus and Neptune because their remote locations in the far outer solar system require transfer times that are a challenge to spacecraft reliability engineering and to radioisotope power systems whose output power decay with time. The VAP strongly influences the $\Delta$$V$ necessary for orbit insertion and the entry speed of an atmospheric entry probe delivered from approach: a higher VAP requires a higher orbit insertion $\Delta$$V$ and thus more of the spacecraft's mass devoted to propellant, and increases the entry speed of the entry probe, requiring a more massive heat shield. The DAP influences the locations available to an entry probe, and influences the probe's atmosphere-relative entry speed because it limits the alignment of the entry velocity vector with the local planetary rotation velocity. Uranus represents an extreme case (in our solar system). Its 97.7$\deg$ obliquity can, over 1/4 of a Uranian orbit ($\sim$21 years), change the average DAP from equator-on to nearly pole-on. These four characteristics are not entirely independent. Trajectories with short transfer durations almost invariably have high VAPs. Trajectories with low VAPs can have high DAPs, especially at Uranus. Mission designers must examine all the options, assessing the interplay of these characteristics and their implications for mission risk, cost, and performance.

Thousands of possible transfer trajectories from Earth to Uranus have been identified, and hundreds to Neptune \citep{Elliott2017}. A few have particularly advantageous combinations of characteristics and are identified as the best options within that study's assumed launch window. Similar, and in some cases better options would be available outside of that study's launch window. For instance, when Jupiter and Saturn align to provide gravity assists from both, trajectories with short transfer durations are possible. Thus, if programmatic considerations dictate a particular launch window, there are useful trajectories available for transfers to either Uranus or Neptune.

\subsubsection{Probe Delivery and Options for Probe Entry Location}
Given a transfer trajectory with its particular VAP and DAP, a remaining degree of freedom, the ``$b$'' parameter (the offset of the $b$-plane aim point from the planet's center), determines both the available entry site locations, and the atmosphere-relative entry speed for each of those locations, and the entry flight path angle (EFPA). If the probe is delivered and supported by a flyby spacecraft, designing a trajectory to give data relay window durations of an hour or more is not difficult. But if the CRSC is an orbiter delivering the probe from hyperbolic approach, the probe mission must compete with the orbit insertion maneuver for performance. Orbit insertion maneuvers are most efficiently done near the planet, saving propellant mass. But such trajectories, coupled with a moderately shallow probe EFPA that keeps entry heating rates and inertial loads relatively low, yield impractically short data relay durations. For this type of trajectory, the orbiter rapidly passes through the probe's data relay antenna beam and the telecommunications time is much shorter. Steepening the entry (decreasing {\it b}) can increase the window duration and requires the CRSC to be on a trajectory with a somewhat more distant closest approach, resulting in a slower overflight and correspondingly increased telecom window, but at the cost of significantly increased entry heating rates and inertial loads. A different approach to this problem, described in the NASA Ice Giants Missions study report, but not analyzed in depth, avoids this situation by delivering the probe to a b-plane aim point $\sim$180$\deg$ away from the orbiter's aim point. Although this requires a minor increase in the orbiter's total $\Delta$V for targeting and divert, it allows a moderate EFPA for the probe while allowing a data relay window of up to two hours duration.

\subsubsection{Ice Giant Entry Challenges}
The probe aeroshell, provided by NASA and NASA Ames Research Center will comprise both a heatshield (foreward aeroshell) and an aft cover (backshell). The aeroshell has five primary functions:

\begin{itemize}[label=\textbullet]
\item To provide an aerodynamically stable configuration during hypersonic and supersonic entry and descent into the H$_2$--He ice-giant atmosphere while spin-stabilized along the probe's symmetry (rotation) axis;
\item To protect the descent vehicle from the extreme heating and thermo-mechanical loads of entry.
\item To accommodate the large deceleration loads from the descent vehicle during hypersonic entry.
\item To provide a safe, stable transition from hypersonic/supersonic to subsonic flight.
\item To safely separate the heatshield and backshell from the descent vehicle based on $g$-switch with timer backup, and transition the descent vehicle to descent science mode beneath the main parachute.
\end{itemize}

One of the primary challenge for an ice-giant probe aeroshell is the heat-shield material and system that can withstand the extreme entry environment. Heritage carbon-phenolic thermal protection system used successfully for the Galileo and Pioneer-Venus entry aeroshell heatshields is no longer feasible due to raw material availability and also processing and manufacturing atrophy. Another challenge is the limitations of ground test facilities needed to requalify a variant of the heritage carbon-phenolic or to develop and certify new material that will ensure survival and function as designed under the extreme entry conditions encountered at the ice giants. Currently, few facilities exist with the necessary capabilities to test thermal performance to the conditions likely to be encountered by an ice-giant probe, including stagnation heat-fluxes between (2.0 kW/cm$^2$--4.0 kW/cm$^2$) and stagnation pressure of 9--12 bars. At Uranus, relative entry velocities are $\sim$22 km/s, and the entry flight path angle determines both the total heat-load and the mechanical (deceleration) load. Steeper entries result in lower total heat-load due to shorter time of flight to reach subsonic velocities but at a significantly higher deceleration (higher $g$-loading), and stagnation heat-flux and pressure. Shallower entries provide lower the $g$-loads and stagnation conditions, but increase the total heat-load. In addition, as mentioned previously -- CRSC trajectories that provide shallower entry flight path angles typically result in the CRSC being much closer to the planet and therefore limit the time available for the probe telecom since the CRSC will pass through the probe antenna beam much more rapidly. All of these constraints, considerations, and trades need to be considered in the probe entry architecture design, and in selecting the TPS materials that can ensure a safe entry. 

\subsubsection{Enabling Technologies}
The need for heat-shield to withstand the extreme entry conditions encountered at the gas giant planet Saturn and the ice giant planets Uranus and Neptune is critical and currently being addressed by NASA. NASA is investing in the development of a new heat-shield material and system technology called Heat-shield for Extreme Entry Environment Technology (HEEET). HEEET will reach TRL6 by 2018. NASA has incentivized and offered HEEET to New Frontiers-4 entry probe mission proposals that are currently under competitive selection considerations.  HEEET, an ablative TPS system that uses 3-D weaving to achieve both robustness and mass efficiency at extreme entry conditions, has been tested at conditions that are relevant for Saturn and Uranus entry probe missions, as well as for missions to Venus and very high-speed sample return missions. Unlike other ablative TPS materials, HEEET is designed to withstand not only extreme entry with a pure carbon recession layer, but is also designed to minimize the heat transferred to the aeroshell structure by having an insulative layer that is much lower density and made of composite material to lower thermal conductivity. These distinct insulative and low thermal conductivity layers not bonded, but are woven together integrally, providing both robustness and efficiency.  Compared to heritage carbon-phenolic system, HEEET is nearly 50\% mass efficient \citep{Ellerby2016}.  

The probe aeroshell will need to be provided by NASA as it is developing and delivering an ablative TPS system to meet the mission needs for extreme entry environments. This allows shallower entry to be considered for entry into an ice giant, Saturn, or Venus.  

There are a number of flight-qualified materials available for backshell TPS. For example, in the backshell the conditions will be typically 2--5\% of the peak stagnation condition on the heat-shield and hence PICA, another NASA developed technology that has been flown at conditions ranging from (100 W/cm$^2$ to 1000 W/cm$^2$) can be used.  The aeroshell design including the 45$\deg$ sphere-cone shape and size proposed for HERA \citep{Mousis2016} will serve as the Uranus aeroshell and the shape is aerodynamically proven at Venus as well as at Jupiter, and will therefore meet the requirements at Uranus. The primary technology challenge for ice giant entry probe missions is the heatshield system and by using HEEET developed by NASA and using NASA expertise, minimal technology development is required.

\subsection{Atmospheric Entry Probe System Design}
\subsubsection{Overview}
The probe comprises two major sub-elements: 1) the descent vehicle including parachutes will carry all the science instruments and support subsystems including telecommunications, power, control, and thermal into the atmosphere, and 2) the aeroshell that protects the descent vehicule during cruise, coast, and entry. The probe (Descent Vehicle + Aeroshell) is released from the CRSC, and arrives at the entry interface point following a long coast period. The Descent Vehicle (including the parachute system) carries the science payload into the deeper atmosphere. It is important to note that although the probe is released from the CRSC and is the vehicle that reaches the entry interface point, and the descent vehicle including parachutes descend into the ice-giant atmosphere, elements of the probe system including the probe release and separation mechanism remain with the CRSC.

Prior to entry, the probe coast timer (loaded prior to probe release) provides a wakeup call to initiate the entry power-on sequence for initial warmup, checks on instrument and subsystem health and status, and pre-entry calibrations. An ice-giant probe can arrive at the entry interface point with an-atmosphere relative velocity in the range of 22--26 km/s. Depending on an entry flight path angle, a probe at Uranus may experience peak heating of 2.5--3.5 kW/cm$^2$, a peak entry deceleration pulse of 165--220 g's, and a stagnation pressure of 9--12 bars. At Neptune, the entry is even more severe with peak heating of 4.3--10 kW/cm$^2$, peak deceleration of 125--455 g's, and stagnation pressures of 7--25 bars \citep{Elliott2017}. The peak heating, total heat soak, and deceleration pulse will depend on the selected mission design including entry location (latitude/longitude), inertial heading, and flight path angle. The probe thermal protection system provides protection for the probe against the intense heating and thermal loads of entry, and an aft cover will protect the back of the probe from somewhat more benign radiative heating environment. 

During descent, the descent vehicule provides a thermally protected environment for the science instruments and probe subsystems, including power, operational command, timing, and control, and reliable telecommunications for returning probe science and engineering data. The probe avionics will collect, buffer, format, process (as necessary), and prepare all science and engineering data to be transmitted to the CRSC. The probe descent subsystem controls the probe descent rate and rotation necessary to achieve the mission science objectives. 

Although the atmospheres of the ice giants have been modeled, the actual thermal, compositional, and dynamical structure beneath the cloud tops is largely unknown. Possible differences in composition and temperature/pressure structure between the atmosphere models and the measured atmosphere have the potential to adversely affect the performance of the probe relay telecom and must be accounted for in selection of communication link frequency. In particular, the microwave opacity of the atmosphere is dependent on the abundances of trace species such as NH$_3$, H$_2$S, and PH$_3$, all microwave absorbers. In general, the opacity of these absorbers increases as the square of the frequency, and this drives the choice of telecom frequency to the lowest frequency reasonable, likely UHF. The final decision on frequency consequently affects the probe transmit antenna design, including structure, size, gain, and beam pattern/beamwidth. Decisions on antenna type and properties also depend on the probe descent science requirements, the time required to reach the target depth, and the CRSC overflight trajectory, including range, range rate, and angle. Throughout descent, the rotation of the planet and the CRSC overflight trajectory, along with atmospheric winds, waves, convection, and turbulence, aerodynamic buffeting, and descent vehicle spin and pendulum motion beneath the parachute will add Doppler contributions to the transmitted frequency that must be tracked by the CRSC receivers.

The ice giants are significantly cooler than the gas giants. At 20 bars, the atmospheres of Jupiter and Saturn reaches about 415 and 355 K, respectively, whereas at Uranus the 10-bar/20-bar temperatures are only about 180/225 K. However, at an altitude of 56 km above 1 bar, the tropopause is an extremely cold: 53 K as compared to the tropopause temperatures on Jupiter and Saturn of 110 and 85 K, respectively. Survival at the low tropospheric temperatures of the ice giants will require careful consideration be given to probe thermal-control design, and may dictate a sealed probe. At Uranus, the 10-bar level is located approximately 160 km beneath tropopause. If the Uranus science goal is to descend to 10 bars within one hour, an average descent speed of 45 m/s is required. With a scale height of about 33 km, a 160 km descent from the tropopause to 10-bars will pass through approximately 5 Uranian scale heights.
			
\subsubsection{Entry Probe Power and Thermal Control}

Following the release of the Descent Vehicle from the CRSC, the descent vehicule has four main functions:

\begin{itemize}[label=\textbullet]
\item To initiate the ``wake up'' sequence at the proper time prior to arrival at the entry interface point.
\item To safely house/protect, provide command and control authority for, provide power for, and maintain a safe thermal environment for all the subsystems and science instruments.
\item To collect, buffer as needed, and relay to the CRSC all required pre-entry, entry, and descent housekeeping, engineering, calibration, and science engineering data. 
\item To control the descent speed and spin rate profile of the descent vehicule to satisfy science objectives and operational requirements. 
\end{itemize}

An ice giant mission will possibly include one or several Venus flybys at 0.7 AU prior to a long cruise to the outer solar system at 20--30 AU. To provide a safe, stable thermal environment for probe subsystems and instruments over this range of heliocentric distances is not a trivial issue, and will require careful thermal design with care given to accounting for and understanding possible heat loss pathways. High-TRL insulating materials, models, and analysis and thermal management techniques will be used in the design program. 

Prior to arrival, the descent vehicule is released from the CRSC for a long coast to the entry interface point. During this coast period, the descent vehicule must maintain safe internal temperatures while providing power for the coast timer and the coast transmitter system needed to provide periodic health checks to the CRSC. While autonomous thermal control can be provided by batteries, an option for replacing the batteries is to add NASA or European Radioisotope Heater Units (RHUs). Since an ice giant flagship mission would almost certainly be nuclear powered, issues related to additional cost and launch approval will have already been addressed. Use of RHUs would significantly reduce the battery complement with significant mass savings likely. Future technology developments with the potential to loosen some of the probe temperature requirements include the development of very low temperature (cryo) electronics.

Once released from the CRSC, the probe will necessarily be entirely self-sufficient for mission operations, thermal control, and power management. As discussed, during coast, safe internal temperatures could be maintained with either RHUs or by way of primary batteries that provide electric power for small heaters as needed. Additional power is needed during coast for the coast timer as well as periodic health and status transmissions to the CRSC. During pre-entry and entry, the batteries support the probe wake-up, turn-on, system health checks and calibration, and entry acceleration measurements and data collection. Under the parachute, the batteries support all probe operations including dual channel data transmission with an RF out of approximately 10 watts/channel. Future technology developments may realize batteries with higher specific energies resulting in potential mass savings. 

\subsubsection{Data Relay}
The probe telecommunication system comprises two redundant channels that, to improve isolation, will transmit orthogonal polarizations at slightly offset frequencies, and will operate in transmit mode only. Once released from the CRSC, the probe can no longer receive any commands. The telecom system is designed to ensure safe and reliable data return from the atmosphere as the probe descends under parachute. Driven by an ultrastable oscillator to ensure a stable link frequency for radio science measurements of atmospheric dynamics, the frequency of the probe to CRSC relay link is chosen primarily based on the microwave absorption properties of the atmosphere. The properties of the Jupiter system that drove the Galileo probe relay link frequency to higher frequencies (L-band) included the intense, pervasive synchrotron radiation from Jupiter's powerful magnetosphere. This is not a significant issue at the ice giants, and due to the increase in microwave opacity with higher frequencies, the relay link operates at UHF frequencies where atmospheric opacity is minimal \citep{Balint2004,Beebe2010}.

The probe data relay includes the transmission of pre-entry and entry engineering and instrument calibration data, measurements of entry accelerations, and all probe descent science acquired by the probe instrument payload. As compared to the single data rate systems utilized by the Galileo \citep{Bright1984} and Huygens \citep{Clausen2002} probes, an ice-giant probe may implement a variable data rate strategy to optimize the data return for the rate at which science data is collected and reflecting the probe descent profile and changing probe-CRSC geometry. The descent sequence and relay link strategy are selected to ensure that all collected science data be successfully transmitted prior to the probe reaching its target depth, nominally 10 bars.

The probe low-gain antenna will be mounted on back of the probe to nominally transmit in the --z direction, opposite to the probe descent velocity vector, and will have a beamwidth large enough to support probe pendulum motion beneath the parachute while allowing for a large range of CRSC zenith angles throughout the probe descent. At UHF frequencies, a microwave patch antenna provides good performance with a peak gain of about 5--6 dB. The probe-relay signal will be received on the CRSC either through a dedicated probe relay antenna, or through the CRSC high gain antenna. Within the CRSC Relay Receiver, radio science data -- frequency and signal strength - is recorded. Since the probe descent science, engineering, and housekeeping data volume is quite small, likely no more than several tens of Mbit, the CRSC is able to store multiple copies of each channel of probe data, with the option available for open loop recording of the probe signal. Following the end of the probe descent mission, the CRSC will return to Earth-point and downlink multiple copies of the stored probe data.

\subsubsection{Carrier Relay Spacecraft}
During the long cruise to the outer solar system, the CRSC provides structural and thermal support, provides power for the probe, and supports periodic health checks, communications for probe science instrument software or calibration changes, and other post-launch software configuration changes and mission sequence loading as might be required from launch to encounter. Upon final approach to Uranus, the CRSC supports a final probe health and configuration check, rotates to the probe release orientation, cuts cables and releases the probe for the probe cruise to the entry interface point. Following probe release, the CRSC may be tracked for a period of time, preferably several days, to characterize the probe release dynamics and improve reconstructions of the probe coast trajectory and entry interface location. An important release sequence option would be to image the probe following release for optical navigation characterization of the release trajectory. Following probe release and once the CRSC tracking period is over, the CRSC is deflected from the planet-impact trajectory required for probe targeting to a trajectory that will properly position the CRSC for receiving the probe descent telecommunications. During coast, the probe will periodically transmit health status reports to the CRSC. Additionally, the CRSC will conduct a planet-imaging campaign to characterize the time evolution of the atmosphere, weather, and clouds at the probe entry site, as well as to provide global context of the entry site. 

Prior to the initiation of the probe descent sequence, the CRSC will rotate to the attitude required for the probe relay receive antenna to view the probe entry/descent location and will prepare to receive both channels of the probe science telecommunications. The CRSC relay-receive antenna could either be a dedicated relay antenna similar to that used on the Galileo orbiter, or the CRSC could use the spacecraft high gain antenna similar to the Cassini-Huygens relay telecommunications configuration. To account for changes in the CRSC antenna pointing due to the trajectory of the CRSC, the rotation of the planet, and the possible effect of winds on the probe descent location, the option for periodic repointing of the CRSC relay receive antenna must be accommodated.

Following receipt of the probe transmission, multiple copies of the entire probe science data set are stored in CRSC memory prior to Earth downlink. It is expected that the memory storage requirements are easily met with a few hundred Mbit of storage capacity. Once the probe mission is completed and all probe data have been relayed to the CRSC, the CRSC will rotate to point the HGA at Earth and, to ensure complete transfer of the entire data set, the CRSC will initiate the first of multiple downlinks of the probe data set.

\subsection{NASA/ESA Collaboration}
The participation of and contributions from NASA are essential for an ESA-led entry probe. The ESA Uranus/Neptune probe mission will begin its flight phase as an element of a NASA Uranus or Neptune mission (likely a NASA Flagship mission) launch to place both the NASA spacecraft, which functions also as the probe's CRSC, and the probe on a transfer trajectory to Uranus or Neptune. The thermal protection necessary to protect the probe during high speed entry is still to be determined, but it is likely to be the HEEET (Heat Shield for Extreme Entry Environment Technology) material currently being developed by NASA. Additionally, NASA may contribute both instruments with Pioneer, Galileo, and Huygens heritage, as well as provide the participation of significant expertise from many engineers and scientists with experience with previous solar system entry probe missions.

\section{Possible Probe Model Payload}
\label{pay}

Table \ref{payload_1} presents a suite of scientific instruments that can address the scientific requirements discussed in previous sections. This list of instruments should be considered as an example of scientific payload that we might wish to see onboard. Ultimately, the payload of a Uranus or Neptune probe would be defined from a detailed mass, power and design trades, but should seek to address the majority of the scientific goals outlined in Sections 2 and 3.

\subsection{Mass Spectrometry}
\label{MS}

The chemical and isotopic composition of Uranus' and Neptune's atmospheres, and their variabilities, will be measured by mass spectrometry. The scientific objectives relevant to the planets' formation and the origin of the solar system requires {\it in situ} measurements of the chemical composition and isotope abundances in the atmosphere, such as H, C, N, S, P, Ge, As, noble gases He, Ne, Ar, Kr, and Xe, and the isotopes D/H, $^{13}$C/$^{12}$C, $^{15}$N/$^{14}$N, $^{17}$O/$^{16}$O, $^{18}$O/$^{16}$O, $^3$He/$^4$He, $^{20}$Ne/$^{22}$Ne, $^{38}$Ar/$^{36}$Ar, $^{36}$Ar/$^{40}$Ar, and those of Kr and Xe, of which very little is known at present (see Sections 2 and 3). At Jupiter, the Galileo Probe Mass Spectrometer (GPMS) experiment \citep{Niemann1992} was designed to measure the chemical and isotopic composition of Jupiter's atmosphere in the pressure range from 0.15 to 20 bar by {\it in situ} sampling of the ambient atmospheric gas. The GPMS consisted of a gas-sampling system that was connected to a quadrupole mass spectrometer. The gas sampling system also had two sample enrichment cells, one for enrichments of hydrocarbons by a factor 100--500, and one for noble gas analysis cell with an enrichment factor of about 10. The abundance of the minor noble gases Ne, Ar, Kr, and Xe were measured by using the enrichment cell on the Galileo mission, but the sensitivity was too low to derive isotope abundances with good accuracy \citep{Niemann1996}. From GPMS measurements the Jupiter He/H$_2$ ratio was determined as 0.1567 $\pm$ 0.006. To improve the accuracy of the measurement of the He/H$_2$ ratio and isotopic ratios by mass spectrometry the use of reference gases will be necessary. The ROSINA experiment on the Rosetta mission carried a gas calibration unit for each mass spectrometer \citep{Balsiger2007}. Similarly, the SAM experiment on the Curiosity rover can use either a gas sample from its on-board calibration cell or utilise one of the six individual metal calibration cups on the sample manipulation system \citep{Mahaffy2012}. 

A major consideration for the mass-spectrometric analysis is how to distinguish between different molecular species with the same nominal mass, e.g., N$_2$,  CO, and C$_2$H$_4$, which all have nominal mass 28, but differ in their actual mass by about 0.01 amu. There are two ways to address this problem, one is high-resolution mass spectrometry with sufficient mass resolution to resolve these isobaric interferences for the molecules of interest (i.e., $m$/$\Delta$$m$ = 3,000 for the given example), and the other way is chemical pre-separation of the sample followed by lower resolution mass spectrometry. 

\subsubsection{High-Resolution Mass Spectrometry}

\begin{figure*}
\begin{center}
\includegraphics[angle=0,width=10cm]{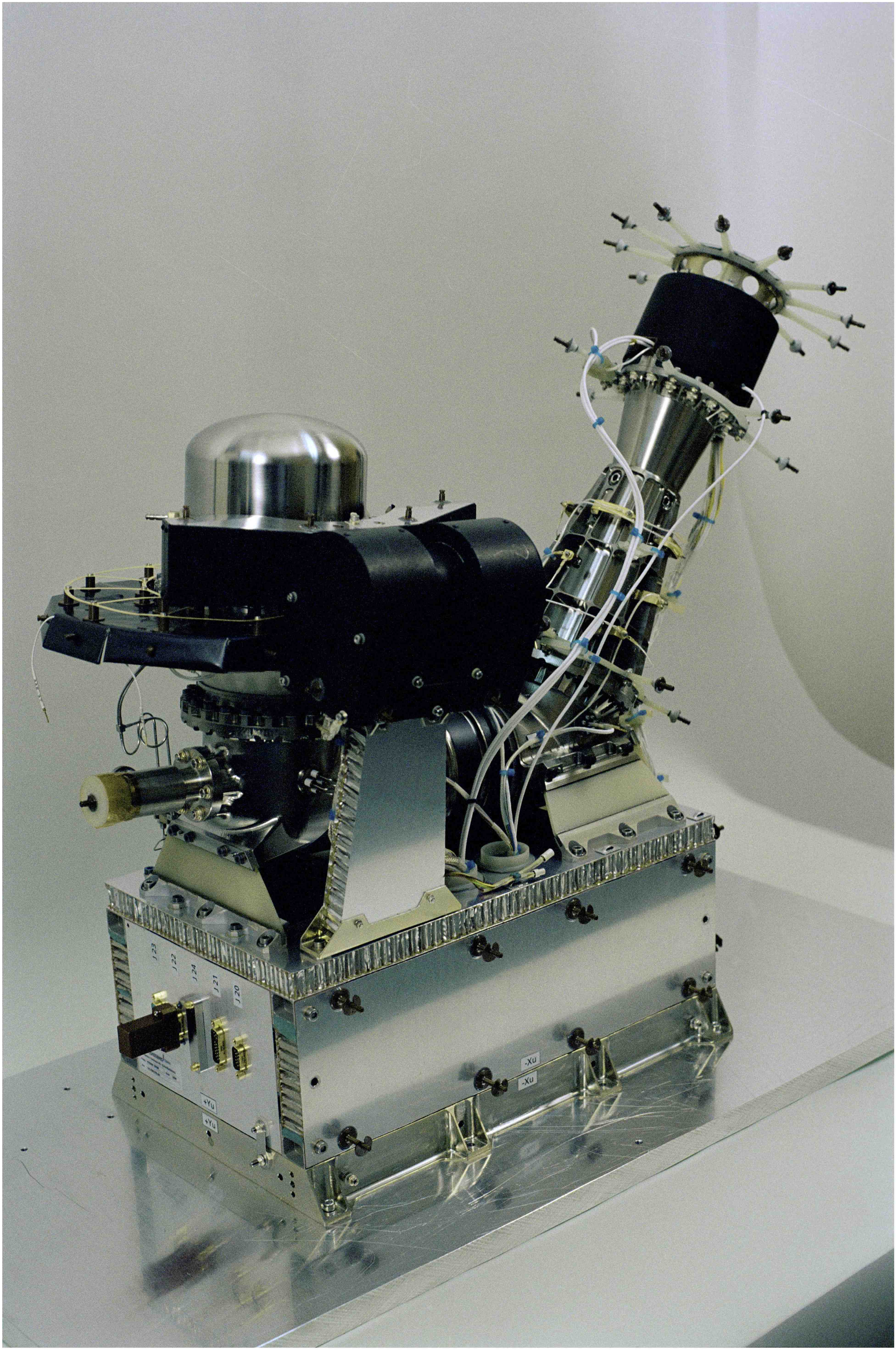}
\end{center}
\caption{Flight model of DFMS/ROSINA instrument without thermal hardware.}
\label{MS_1} 
\end{figure*}

High-resolution mass spectrometry is defined by the capability of the mass spectrometer to resolve isobaric interferences. Usually that means mass resolution of 10,000 and larger, depending on the nature of the isobaric interference. Probably the first high-resolution mass spectrometer in space is the ROSINA experiment on the Rosetta mission \citep{Balsiger2007}. ROSINA has a Double-Focussing Mass Spectrometer (DFMS), see Figure \ref{MS_1}, with a mass resolution of about $m$/$\Delta$$m$ = 9,000 at 50 percent peak height (corresponding to $m$/$\Delta$$m$ = 3,000 at 1\% peak height), Reflectron-Time-of-flight (RTOF) instrument with a mass resolution of about $m$/$\Delta$$m$ = 5,000 at 50\% peak height \citep{Scherer2006}, and a pressure gauge. Determination of isotope ratios with an accuracy at the percent-level has been accomplished for gases in the cometary coma for H/D \citep{Altwegg2015}, for $^{12}$C/$^{13}$C and $^{16}$O/$^{18}$O \citep{Hassig2017}, for $^{35}$Cl/$^{37}$Cl and $^{79}$Br/$^{81}$Br \citep{Dhooghe2017}, for the silicon isotopes \citep{Rubin2017}, $^{36}$Ar/$^{38}$Ar \citep{Balsiger2015}, and Xe isotopes \citep{Marty2017}.

A time-of-flight instrument with even more mass resolution has been developed for possible application in Europa's atmosphere, which uses a multi-pass time-of flight configuration \citep{Brockwell2016}. Accomplished mass resolutions are $m$/$\Delta$$m$ = 40,000 at 50\% peak height and 20,000 at 10\% peak height. An alternative multi-pass time-of-flight instrument has been developed by \cite{Okumura2004}, which uses electric sectors instead of ion mirrors for time and space focussing, which allows for high mass resolution in a compact design. Mass resolutions up to $m$/$\Delta$$m$ = 350,000 have been reported \citep{Toyoda2003}. Later, a more compact version of this instrument has been developed \citep{Shimma2010,Nagao2014}. 

Recently, a new type of mass spectrometer, the Orbitrap mass spectrometer, was introduced \citep{Makarov2000,Hu2005}, which uses ion confinement in a harmonic electrostatic potential. The Orbitrap mass spectrometer is a Fourier-Transform type mass spectrometer, and it allows for very high mass resolutions in a compact package. Resolving powers above 1,000,000 have been accomplished with laboratory instruments \citep{Denisov2012}. For example, using an Orbitrap mass spectrometer for laboratory studies of chemical processes in Titan's atmosphere, mass resolutions of $m$/$\Delta$$m$ = 100,000 have been accomplished up to $m$/$z$ = 400 \citep{Horst2012}, and $m$/$\Delta$$m$ = 190,000 at 50\% peak height and $m$/$z$ = 56 in a prototype instrument for the JUICE mission \citep{Briois2013,Briois2016}. 

\subsubsection{Low-Resolution Mass Spectrometry with Chemical Pre-processing}

The alternative approach to high-resolution mass spectrometry, is to use a simpler low-resolution mass spectrometer together with a chemical processing of the sample to separate or eliminate isobaric interferences. One established way used in space instrumentation is to use chromatographic columns with dedicated chemical specificity for a separation of chemical substances. Also enrichments cells to selectively collect a group of chemical species have been used. 

The Gas-Chromatograph Mass Spectrometer (GCMS) of the Huygens probe is a good example of such an instrument \citep{Niemann2002,Niemann2005,Niemann2010}. The Huygens probe GCMS has three chromatographic columns, one column for separation of CO and N$_2$ and other stable gases, the second column for separation of nitriles and other organics with up to three carbon atoms, and the third column for the separation of C$_3$ through C$_8$ saturated and unsaturated hydrocarbons and nitriles of up to C$_4$. The GCMS was also equipped with a chemical scrubber cell for noble gas analysis and a sample enrichment cell for selective measurement of high boiling point carbon containing constituents. A quadrupole mass spectrometer was used for mass analysis with a mass range from 2 to 141 u/e, which is able to measure isotope ratios with an accuracy of 1\%. 

Examples of newer GCMS instrumentation are the Ptolemy instrument on the Rosetta lander for the measurement of stable isotopes of key elements \citep{Wright2007}, which uses an ion trap mass spectrometer, the COSAC instrument also on the Rosetta lander for the characterisation of surface and subsurface samples \citep{Goesmann2007}, which uses a time-of-flight mass spectrometer, the GCMS instrument for the Luna-Resource lander \citep{Hofer2015}, which also uses a time-of-flight mass spectrometer, and the SAM experiment on the Curiosity rover \citep{Mahaffy2012}, which uses a classical quadrupole mass spectrometer.

To increase the sensitivity for a range of chemical compounds (e.g. hydrocarbons) dedicated enrichment cells were used, as discussed above for the GPMS experiment. A novel and promising enrichment cell uses the cryotrapping technique, which has a long history in the laboratory. The use of cryotrapping increases the instruments sensitivity by up to 10,000 times the ambient performance \citep{Brockwell2016}, and would allow for the detection of noble gases at abundances as low as 0.02 ppb \citep{Waite2014}.

\subsubsection{Summary of Mass Spectrometry}

So far in most space missions the chemical pre-separation was the technique used to overcome isobaric interferences in the mass spectra, with the exception of the mass spectrometer experiment ROSINA on the Rosetta orbiter. Chemical pre-separation works well, but by choosing chromatographic columns with a certain chemical specificity one makes a pre-selection of the species to be investigated in detail. This is a limitation when exploring an object of which little is known. Also, gas chromatographic systems with several columns are rather complex systems, both to build and to operate (see the SAM instrument as a state-of-the art example of this technique; \cite{Mahaffy2012}).

In recent years there has been a significant development of compact mass spectrometers that offer high mass resolution. Thus, solving the problem of isobaric interferences in the mass spectra by mass resolution can be addressed by mass spectrometry alone and one should seriously consider using high-resolution mass spectrometry for a future mission to probe planetary atmospheres. After all, no a priori knowledge of the chemical composition has to be assumed in this case. In addition, with modern time-of-flight mass spectrometers mass ranges beyond 1000 u/e are not a problem at all, which, for example, would have been useful to investigate Titan's atmosphere. Nevertheless, enrichments of certain chemical groups (e.g., hydrocarbons or noble gases) should still be considered even in combination with high-resolution mass spectrometry to maximise the science return. 

\subsubsection{Tunable Laser System}

A Tunable Laser Spectrometer (TLS) \citep{Durry2002} can be employed as part of a Gas-Chromatograph system to measure the isotopic ratios to a high accuracy of specific molecules, e.g. H$_2$O, NH$_3$, CH$_4$, CO$_2$ and others. TLS employs ultra-high spectral resolution (0.0005 cm$^{-1}$) tunable laser absorption spectroscopy in the near infra-red (IR) to mid-IR spectral region. TLS is a simple technique that for small mass and volume can produce remarkable sensitivities at the sub-ppb level for gas detection. Species abundances can be measured with accuracies of a few \%. With a TLS system one can derive isotope abundances with accuracies of about 0.1\% for the isotopic ratios of D/H, $^{13}$C/$^{12}$C, $^{18}$O/$^{16}$O, and $^{17}$O/$^{16}$O. 

For example, TLS was developed for application in the Mars atmosphere \citep{Lebarbu2004}, within the ExoMars mission; a recent implementation of a TLS system was for the Phobos Grunt mission \citep{Durry2010}, and another TLS is part of the SAM instrument on the Curiosity Rover \citep{Webster2011}, which was used to measure the isotopic ratios of D/H and of $^{18}$O/$^{16}$O in water and $^{13}$C/$^{12}$C, $^{18}$O/$^{16}$O, $^{17}$O/$^{16}$O, and $^{13}$C$^{18}$O/$^{12}$C$^{16}$O in carbon dioxide in the Martian atmosphere \citep{Webster2013}.

\subsection{Helium Abundance Detector}

The Helium Abundance Detector (HAD), as it was used on the Galileo mission \citep{vonZahn1992}, measures the refractive index of the atmosphere in the pressure range of 2--10 bar. The refractive index is a function of the composition of the sampled gas, and since the jovian atmosphere consists of mostly of H$_2$ and He, to more than 99.5\%, the refractive index is a direct measure of the He/H$_2$ ratio. The refractive index can be measured by any two-beam interferometer, where one beam passes through a reference gas and the other beam through atmospheric gas. The difference in the optical path gives the difference in refractive index between the reference and atmospheric gas. For the Galileo mission, a Jamin-Mascart interferometer was used, because of its simple and compact design, with an expected accuracy of the He/H$_2$ ratio of $\pm$0.0015. The accomplished measurement of the He mole fraction gave 0.1350 $\pm$ 0.0027 \citep{vonZahn1998}, with a somewhat lower accuracy than expected, but still better than is possible by a mass spectrometric measurement.

\subsection{Atmospheric Structure Instrument}
\label{ASI}

The Atmospheric Structure Instrument (ASI) of the entry probe will make {\it in situ} measurements during the entry and descent into the atmosphere of Uranus and Neptune in order to investigate the atmospheric structure, dynamics and electricity. The scientific objectives for ASI are to determine the atmospheric profiles of density, pressure and temperature along the probe trajectory and the investigation of the atmospheric electricity (e.g. lightning) by {\it in situ} measurements. The ASI will use the mean molecular weight as measured by the mass spectrometer to calculate the profile of atmospheric density. 

The ASI benefits from the strong heritage of the Huygens ASI experiment of the Cassini/Huygens mission \citep{Fulchignoni2002} and Galileo, and Pioneer Venus ASI instruments \citep{Seiff1992,Seiff1980}. The key {\it in situ} measurements will be entry accelerations from which the density of the upper atmosphere (above parachute deployment) can be found, and from this the pressure and temperature profiles can be retrieved. During parachute descent, the ASI will perform direct temperature and pressure sampling \citep{Fulchignoni2005,Seiff1998}. Once the probe heat shield is jettisoned, direct measurements of pressure, temperature and electrical properties will be performed. During descent, the pressure, temperature, and and electric property sensors will be placed beyond the probe boundary layer to have unimpeded access to the atmospheric flow.

{\it In situ} measurements are essential for the investigation of the atmospheric structure and dynamics. The data provided by the ASI will help constrain and validate models of atmospheric thermal, electrical, and dynamical structure. The ASI measurement of the atmospheric pressure and temperature will constrain the stability of the atmosphere, providing an important context for understanding the atmospheric dynamics and mixing and the energy and cloud structure of the atmosphere. The determination of the lapse rate can help identify locations of condensation and eventually clouds, and to distinguish between saturated and unsaturated, stable and conditionally stable regions. The possible variations atmospheric stability and detection of atmospheric stratification are strongly correlated with the presence of winds, thermal tides, waves, and turbulence within the atmosphere.

The ASI will measure properties of Uranus and Neptune's atmospheric electricity by determining the conductivity profile of the troposphere, and detecting the atmospheric DC electric field. These measurements provide indirect information about galactic cosmic ray ionization, aerosol charging inside and outside of clouds, properties of potential Schumann resonances, and allow for detection of possible electrical discharges (i.e. lightning). ASI could measure the unknown lightning spectra in the frequency range of $\sim$1--200 kHz below the ionosphere, and will obtain burst waveforms with different temporal resolutions and durations in order to detect and characterize lighting activity in ice giants. Refining the location of lightning flashes, whether determined optically from an orbiter or {\it in situ} from a probe, and correlating the detected lightning with the observations of weather systems may provide powerful constraints on the location of deep storms and weather systems and the depth, location, and density of clouds.

\subsection{Doppler-Wind Experiment}

The probe Uranus/Neptune Radio Science Experiment (RSE) will include a Doppler Wind Experiment (DWE) dedicated to the measurement of the vertical profile of the zonal (east-west) winds along the probe descent path, and a measurement of the integrated atmospheric microwave absorption measurements along the probe-relay atmospheric raypath. The absorption measurement will indirectly provide a measurement of atmospheric abundance of ammonia. This technique was used by the Galileo probe to constrain the Jovian atmospheric NH$_3$ profile, strongly complementing measurements of the atmospheric composition by the probe Mass Spectrometer \citep{Folkner1998}. 

The primary objectives of the probe Doppler Wind Experiment is to use the probe-CRSC radio subsystem (with elements mounted on both the probe and the Carrier) to measure the altitude profile of zonal winds along the probe descent path under the assumption that the probe in terminal descent beneath the parachute will accurately trace the zonal wind profile. In addition to the vertical profile of the zonal winds, the DWE will also be sensitive to atmospheric turbulence, aerodynamic buffeting, and atmospheric convection and waves that disrupt the probe descent speed. Key to the Doppler wind measurement is an accurate knowledge of the reconstructed probe location at the beginning of descent, the reconstructed probe descent speed with respect to time/altitude, and the reconstructed Carrier position and velocity throughout the period of the relay link. The probe entry trajectory reconstruction from the entry interface point to the location of parachute deployment depends on measured accelerations during entry, and the descent profile is reconstructed from measurements of pressure and temperature by the Atmospheric Structure Instrument. From the known positions and velocities of the descent probe and Carrier, a profile of the expected relay link frequency can be created, and when differenced with the measured frequencies, a profile of Doppler residuals results. Inversion of the Doppler residual profile using an algorithm similar to the Galileo probe Doppler Wind measurement \citep{Atkinson1997,Atkinson1998}. To generate the stable probe relay signal, the probe will carry a quartz crystal ultrastable oscillator (USO) within the relay transmitter, with an identical USO in the relay receiver on the Carrier spacecraft.

Secondary objectives of the DWE include the analysis of Doppler modulations and frequency residuals to detect, locate, and characterize regions of atmospheric turbulence, convection, wind shear, and to provide evidence for and characterize atmospheric waves. Analysis of the relay link signal strength measurements be used to study the effect of refractive-index fluctuations in Uranus's atmosphere including scintillations and atmospheric turbulence \citep{Atkinson1998,Folkner1998}.

\subsection{Nephelometer}
A nephelometer will be used to characterize the atmospheric clouds, aerosols and condensates. Measurement of scattered visible light within the atmosphere is a powerful tool to retrieve number density and size distribution of liquid and solid particles, relied to their formation process, and to understand the overall character of the atmospheric aerosols based on their refractive index (liquid particles, iced particles, solid particles from transparent to strongly absorbing, etc.). In general, counting instruments are performing their measurements at a given scattering angle, typically around 90$\deg$, considering the scattering properties of the particles that cross a laser beam. The particle concentrations are retrieved in several size classes typically between few tenths of $\mu$m to several tens of $\mu$m \citep{Grimm2009}. The scattered light is dependent both on the size of the particles and the complex refractive index. To accurately retrieve the size distribution, the nephelometer must be calibrated assuming one nature of particles. Typically, carbonaceous particles could be tens of times less luminous than liquid droplets. On the other hand, measurements at small scattering angle below 20$\deg$ are less dependent on the refractive index and can be used for the determining number densities of the aerosols independent of their nature \citep{Renard2010,Lurton2014}.

The retrieval of the full scattering function by a nephelometer that simultaneously records scattered light at different angles by all the particles in the field of view can provide a good estimate of the nature of the particles, particularly refractive index. The size distribution (expected to be monomodal) can be retrieved using Mie scattering theory or more sophisticated models for regular particles having symmetries \citep{Verhaege2009}. Ray tracing method can also be used for large particles as ice crystal \citep{Shcherbakov2006}. It is also possible to distinguish between liquid droplets and iced particles, as done in the Earth atmosphere \citep{Gayet1997}. In the case of irregular shaped particles, the observed scattering function can be compared to reference measurements obtained in laboratory \citep{Renard2002,Volten2006} to identify their nature; the laboratory scattering functions were obtained for a cloud of levitating particles with well-known size distribution.

Due to the low temperature, ice particles of methane and other hydrocarbons are present in the atmospheres of Uranus and Neptune \citep{Sanchez2004,Sanchez2011}. It is then necessary to be able to distinguish between solid and liquid particles when performing light-scattering measurements inside these atmospheres. It is proposed to use the ``LOAC (Light Optical Aerosol Counter)'' concept, already used in routine for {\it in situ} measurements inside the Earth atmosphere \citep{Renard2016a,Renard2016b}, to retrieve both the size distribution in 20 size classes and the scattering function to identify the nature of the particles. At present, LOAC performs measurement at two scattering angles, around 15$\deg$ and 60$\deg$. Scattering at the smaller angle is used to retrieve the size distribution, and scattering at the larger angle combined with smaller angle scattering provides an estimate of the main nature of the aerosols, whether liquid droplets, mineral particles, carbonaceous particles, ice particles, etc. The nature estimate is based on a comparison with laboratory data of the size evolution of the 60$\deg$-angle measurements. To be able to estimate more accurately the nature of the particles for all the size classes in the 0.1--100 $\mu$m size range, measurement must be conducted simultaneously by a ring of 10 to 15 detectors in the 10$\deg$--170$\deg$ scattering angle range. These measurements can be compared to theoretical calculation for droplets and ices, but also to laboratory measurements in case of more complex particles both in shape and in composition.

LOAC used in Earth atmosphere has a pump to inject the particles inside the optical chamber and the laser beam. In case of an atmospheric descent probe, a collecting inlet can be mounted in front of the pump, to inject directly the particles inside the chamber without the pumping system. A dedicated fast electronic will be developped to be able to record accurately the light pulse when particles will cross one by one a thin laser beam at a speed of several tens of m/s, and to be able to detect up to 1000 particles per cm$^3$.

\subsection{Net Energy Flux Radiometer}
\subsubsection{Scientific Impetus}

Ice giant meteorology regimes depend on internal heat flux levels. Downwelling solar insolation and upwelling thermal energy from the planetary interior can have altitude and location dependent variations. Such radiative-energy differences cause atmospheric heating and cooling, and result in buoyancy differences that are the primary driving force for Uranus and Neptune's atmospheric motions \citep{Allison1991,Bishop1995}. The three-dimensional, planetary-scale circulation pattern, as well as smaller-scale storms and convection, are the primary mechanisms for energy and mass transport in the ice giant atmospheres, and are important for understanding planetary structure and evolution \citep{Lissauer2005,Dodson2010,Turrini2014}. These processes couple different vertical regions of the atmosphere, and must be understood to infer properties of the deeper atmosphere and cloud decks (see Figure \ref{clouds}). It is not known in detail how the energy inputs to the atmosphere interact to create the planetary-scale patterns seen on these ice giants \citep{Hofstadter2017}. Knowledge of net vertical energy fluxes would supply critical information to improve our understanding of atmospheric dynamics.

A Net Flux Radiometer (NFR) will contribute to this understanding by measuring the up- and down-welling radiation flux, $F$, as a function of altitude. The net flux, the difference between upward and downward radiative power per unit area crossing a horizontal surface per unit area is directly related to the radiative heating or cooling of the local atmosphere. At any point in the atmosphere, radiative power absorbed per unit volume is given by the vertical derivative of net flux ($dF/dz$) in the plane-parallel approximation where the flux is horizontally uniform; the corresponding heating rate is then ($dF/dz$)/($\rho C_p$), where $\rho$ is the local atmospheric density and $C_p$ is the local atmospheric specific heat at constant pressure.

\subsubsection{Measuring Net Energy Flux}

Three NFR instruments have flown to planets in the past, namely the large probe infrared radiometer \citep{Boese1980} on Pioneer-Venus large probe, small probe NFR on Pioneer-Venus small probe \citep{Colin1977}, and the NFR on the jovian Galileo probe \citep{Sromovsky1992} for {\it in situ} measurements within the venusian and jovian atmospheres, respectively. These instruments were designed to measure the downward and upward radiation flux within their respective atmospheres as the probe descended by parachute. The Galileo NFR encountered rapid temperature excursions during the drop \citep{Sromovsky1998}, a fact that influences the design of the next-generation NFR. The Galileo NFR also measured the vertical profile of upward and downward radiation fluxes on Jupiter from about 0.44 to 14 bars \citep{Sromovsky1998}. Radiation was measured in five broad spectral bands, 0.3--3.5 $\mu$m (total solar radiation), 0.6--3.5 $\mu$m (total solar radiation weighted to the methane absorption region), 3--500 $\mu$m (deposition and loss of thermal radiation), 3.5--5.8 $\mu$m (window region with low gas phase absorption), and 14--35 $\mu$m (hydrogen dominated). Galileo NFR data provided signatures of ammonia (NH$_3$) ice clouds and ammonium hydrosulfide (NH$_4$SH) clouds \citep{Sromovsky1998}. The water fraction was found to be much lower than solar and no water clouds were identified. 

For Uranus and Neptune, the thermal structure and the nominal NFR measurement regime extends from $\sim$0.1 bar (near the tropopause which coincides with the temperature minimum) to at least 50 bar (see Figure \ref{clouds}), the uppermost cloud layer at $\sim$1 bar level is made up of CH$_4$ ice (revealed by Voyager-2 radio occultation observations). The base of the water-ice cloud for solar O/H is expected to be at $\sim$40-bar level, whereas for the NH$_3$-H$_2$O solution clouds $\sim$80 bar \citep{Atreya2004}. So far, only an upper limit is known for Uranus' heat flow based on Voyager 2 \citep{Pearl1990}. {\it In situ} probe measurements will help to define sources and sinks of planetary radiation, regions of solar energy deposition, and provide constraints on atmospheric composition and cloud layers. Ultimately, an NFR in concert with a suite of additional science instruments (mass spectrometer, atmospheric structure suite, nephelometer, radio science /Doppler wind instrument, {\it etc}.) will constrain the processes responsible for the formation of these ice giants. 

\subsubsection{Basic Design Considerations}

\begin{figure*}
\begin{center}
\resizebox{\hsize}{!}{\includegraphics[angle=0]{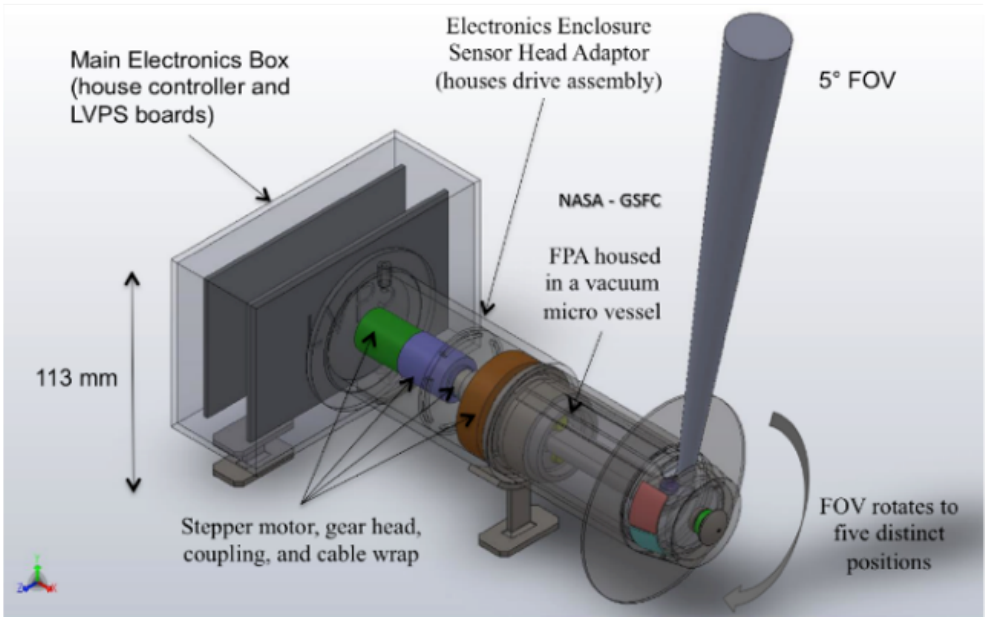}}
%\resizebox{\hsize}{!}{\includegraphics[angle=0]{Saturn_Kzz1d9.pdf}}
\end{center}
\caption{NASA/GSFC NFR instrument concept showing a 5$\deg$ field-of-view that can be rotated by a stepper motor into five distinct look angles. }
\label{NFR_2} 
\end{figure*}

\begin{figure*}
\begin{center}
\resizebox{\hsize}{!}{\includegraphics[angle=0]{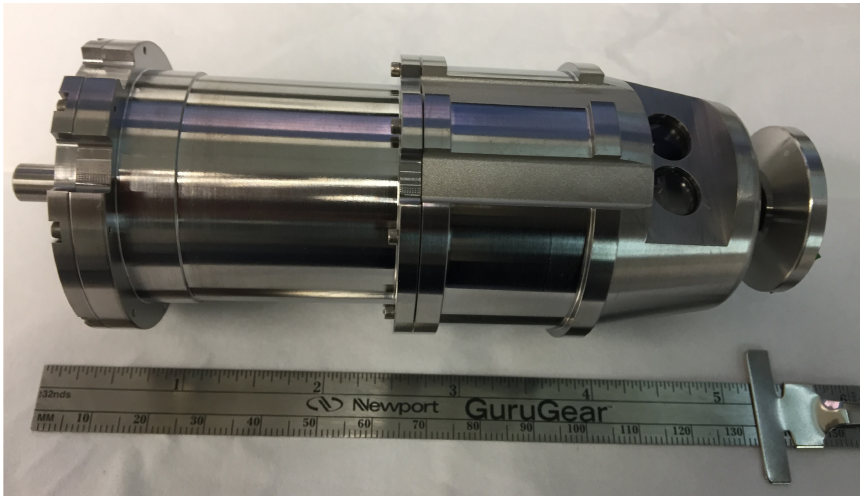}}
%\resizebox{\hsize}{!}{\includegraphics[angle=0]{Saturn_Kzz1d9.pdf}}
\end{center}
\caption{Saturn probe prototype NFR vacuum micro-vessel with sapphire and diamond windows; this houses a focal plane assembly that accommodates Winston cones with a 5$\deg$ field-of-view acceptance angle.}
\label{NFR_3} 
\end{figure*}

Since the days of the Galileo probe NFR, there have been substantial advancements in optical windows and filters, uncooled thermal detectors, and radiation hard electronic readout technologies that have enabled the development of a more capable NFR. The Saturn probe prototype NFR (see Table \ref{payload_1} and Figures \ref{NFR_2} and \ref{NFR_3}) developed at NASA Goddard Space Flight Center \citep{Aslam2015} is designed to measure radiation flux in a 5$\deg$ field-of-view based on the planetary scale height, in two spectral channels (i.e., a solar channel between 0.25 to 5 $\mu$m and a thermal channel between 4 to 50 $\mu$m). The radiometer is capable of viewing five distinct look angles ($\pm$80$\deg$, $\pm$45$\deg$, and 0$\deg$) into the atmosphere during the probe descent. Non-imaging Winston cones with window and bandpass filter combinations define the spectral channels with a 5$\deg$ Field-Of-View (FOV); if necessary and appropriate relaxing the FOV to $>$5$\deg$ is easily implemented, with the added benefit of a smaller focal plane package due to smaller Winston cones. Uncooled single-pixel thermopile detectors are used in each spectral channel and are read out using a custom designed Multi-Channel Digitizer (MCD) Application Specific Integrated Circuit (ASIC) \citep{Aslam2012,Quilligan2015,Quilligan2014}.

For applications to Uranus or Neptune, the solar channel would be essentially preserved, and the thermal channel range extended to capture the majority of the thermal radiation, as the planetary Planck function peak moves to longer wavelengths with colder temperatures and addition of several judiciously chosen and optimized spectral channels (up to seven, with hexagonal close packing of Winston cones, see Sec. \ref{OFC}) to capture radiation flux of gases and particulates and thus provide important independent constraints of atmospheric composition, cloud structure, and scattering processes.

\subsubsection{Optimal Filter Channels}
\label{OFC}

Voyager-2 radio occultation data \citep{Lindal1987} from Uranus for example shows that C is enhanced by more than an order of magnitude with respect to solar abundance. If the mixing ratios of O, S, N, and C are in relative solar abundance then thermochemical equilibrium models \citep{Atreya2004,West1990}, predict that a water cloud will form at deep levels ($>$100 bar), an NH$_4$SH cloud will form at a few tens of bars pressure, NH$_3$ ice will condense near the 10-bar level, and CH$_4$ ice will condense near the 1 bar level. To date the gross features of the upper atmosphere as predicted by these models remain valid but fundamental questions still remain i.e., what levels of solubility of NH$_3$ and CH$_4$ will lead to appreciable depletions in the mixing ratios of these constituents above the water cloud? Also it is not clear that the relative mixing ratios of O, S, N and C are close to solar ratios \citep{Cavalie2017}, since almost all of the enhanced abundances of these elements are due to preferential accumulation of planetesimals (as opposed to gas) by the giant planets and to the partial dissolution of these solid bodies in the forming planets' gaseous envelopes \citep{Pollack1986}. An enhancement of the S to N ratio could deplete NH$_3$ in the upper atmosphere by promoting NH$_4$SH to the point where no NH$_3$ clouds form, but rather an H$_2$S ice cloud may form near the 100 K temperature level where the pressure is about 2 bar. To address these important science questions, contribution functions have been calculated (i.e., the altitude sensitivity of the planet's emergent radiance) for specific infrared channels to demonstrate that an optimal set of filters will be able to probe the methane cloud opacity and tropospheric temperatures from the cloud tops to the tropopause. Seven NFR baseline spectral filter channels, (see Table \ref{payload_3}), have been identified, suitable for both Uranus and Neptune, to probe tropospheric aerosol opacity in the cloud-forming region using dedicated channels near 5 and 8.6 $\mu$m, plus far-infrared channels long ward of 50 $\mu$m and in the visible.

NFR measurements in concert with mass spectrometry of a host of chemical species from cloud-forming volatiles and disequilibrium species tracing tropospheric dynamics will ultimately aid in understanding middle atmospheric chemistry and circulation and cloud-condensation microphysics of the cloud decks.  

\section{Conclusions}
\label{conc}

The next great planetary exploration mission may well be a flagship mission to one of the ice giant planets, possibly Uranus with its unique obliquity and correspondingly extreme planetary seasons, its unusual dearth of cloud features and radiated internal energy, a tenuous ring system and multitude of small moons, or to the Neptune system, with its enormous winds, system of ring arcs, sporadic atmospheric features, and large retrograde moon Triton, possibly a captured dwarf planet. Following previous explorations of the terrestrial planets and the gas giants, the ice giant planets represent the last unexplored class of planets in the solar system. Extended studies of one or both ice giants, including {\it in situ} with an entry probe, are necessary to further constrain models of solar system formation and chemical, thermal, and dynamical evolution, the atmospheric formation, evolution, and processes, and to provide additional groundtruth for improved understanding of extrasolar planetary systems. The giant planets, gas and ice giants together, additionally offer a laboratory for studying the dynamics, chemistry, and processes of Earth's atmosphere. Only {\it in situ} exploration by a descent probe (or probes) can unlock the secrets of the deep, well-mixed atmospheres where pristine materials from the epoch of solar system formation can be found. Particularly important are the noble gases, undetectable by any means other than direct sampling, that carry many of the secrets of giant planet origin and evolution. Both absolute as well as relative abundances of the noble gases are needed to understand the properties of the interplanetary medium at the location and epoch of solar system formation, the delivery of heavy elements to the ice giant atmospheres, and to help decipher evidence of possible giant planet migration. A key result from a Uranus or Neptune entry probe would be the indication as to whether the enhancement of the heavier noble gases found by the Galileo probe at Jupiter (and hopefully confirmed by a future Saturn probe) is a feature common to all the giant planets, or is limited only to the gas giants. 

The primary goal of an ice-giant entry-probe mission is to measure the well-mixed abundances of the noble gases He, Ne, Ar, Kr, Xe and their isotopes, the heavier elements C, N, S, and P, key isotope ratios $^{15}$N/$^{14}$N, $^{13}$C/$^{12}$C, $^{17}$O/$^{16}$O and $^{18}$O/$^{16}$O, and D/H, and disequilibrium species CO and PH$_3$ which act as tracers of internal processes, and can be achieved by an ice-giant probe reaching 10 bars. In addition to measurements of the noble gas, chemical, and isotopic abundances in the atmosphere, a probe would measure many of the chemical and dynamical processes within the upper atmosphere, providing an improved context for understanding the ice giants, the entire family of giant planets (gas giants and ice giants), and the solar system, and to provide ground-truth measurement to improve understanding of extrasolar planets. A descent probe would sample atmospheric regions far below those accessible to remote sensing, well into the cloud forming regions of the troposphere to depths where many cosmogenically important and abundant species are expected to be well-mixed. Along the probe descent, the probe would provide direct tracking of the planet's atmospheric dynamics including zonal winds, waves, convection and turbulence, measurements of the thermal profile and stability of the atmosphere, and the location, density, and composition of the upper cloud layers. 

Results obtained from an ice-giant probe are necessary to improve our understanding of the processes by which the ice giants formed, including the composition and properties of the local solar nebula at the time and location of ice giant formation. By extending the legacy of the Galileo probe mission and possibly a future Saturn entry probe mission, Uranus and Neptune probe(s) would further discriminate between and refine theories addressing the formation, and chemical, dynamical, and thermal evolution of the giant planets, the entire solar system including Earth and the other terrestrial planets, and the formation of other planetary systems.

\section*{Acknowledgements}

The work contributed by O.M., B.B. and T.R. was carried out thanks to the support of the A*MIDEX project (n\textsuperscript{o} ANR-11-IDEX-0001-02) funded by the ``Investissements d'Avenir'' French Government program, managed by the French National Research Agency (ANR). We acknowledge support from the ``Institut National des Sciences de l'Univers''  (INSU), the ``Centre National de la Recherche Scientifique'' (CNRS) and ``Centre National d'Etude Spatiale'' (CNES). Parts of this research were carried out at the Jet Propulsion Laboratory, California Institute of Technology, under a contract with the National Aeronautics and Space Administration. D.H.A, M.D.H., G.S.O., K.R. and C.S. were supported by NASA funds to the Jet Propulsion Laboratory, California Institute of Technology. L.N.F was supported by a Royal Society Research Fellowship and European Research Council Grant at the University of Leicester.  R.H. and A.S.L. were supported by the Spanish MINECO project AYA2015-65041-P (MINECO/FEDER, UE) and Grupos Gobierno Vasco IT-765-13. P.W. acknowledges support from the Swiss National Science Foundation. J.H.W. acknowledges the support of Southwest Research Institute.

\clearpage

\clearpage

\clearpage

\clearpage

\clearpage

\clearpage

\clearpage

%\clearpage

%\begin{figure*}
%\begin{center}
%\resizebox{\hsize}{!}{\includegraphics[angle=0,width=10cm]{NFR_1.pdf}}
%\end{center}
%\caption{The calculated thermal structure of Uranus and Neptune's atmosphere below the lower stratosphere. The predicted cloud decks for Uranus are shown.}
%\label{NFR_1} 
%\end{figure*}

\clearpage

\clearpage

\clearpage

\begin{table*}
\begin{center}
\caption[]{Elemental abundances in Jupiter, Saturn, Uranus and Neptune, as derived from upper tropospheric composition \label{table1}}
\begin{tabular}{lllll}
\hline
\noalign{\smallskip}
Elements	& Jupiter 								& Saturn						& Uranus						& Neptune 										\\
\hline
\noalign{\smallskip}
He/H	$^{(1)}$ 		& ($7.85 \pm 0.16) \times 10^{-2}$		& $(6.75 \pm 1.25) \times 10^{-2}$		&  $(8.88 \pm 2.00) \times 10^{-2}$	& $(8.96 \pm 1.46) \times 10^{-2}$		\\
Ne/H$^{(2)}$ 		& ($1.240 \pm 0.014) \times 10^{-5}$		& --								&  --							& --										\\
Ar/H$^{(3)}$ 		& ($9.10 \pm 1.80) \times 10^{-6}$		& --								& --							& --										\\
Kr/H$^{(4)}$ 		& ($4.65 \pm 0.85) \times 10^{-9}$		& --								& --							& --										\\
Xe/H	$^{(5)}$	 	& ($4.45 \pm 0.85) \times 10^{-10}$		& --								& --							& --										\\
C/H$^{(6)}$ 		& ($1.19 \pm 0.29) \times 10^{-3}$		& $(2.65 \pm 0.10) \times 10^{-3}$		& $(0.6-3.2) \times 10^{-2}$ 		& $(0.6-3.2) \times 10^{-2}$					\\
N/H$^{(7)}$ 		& ($3.32 \pm 1.27) \times 10^{-4}$		& $(0.50-2.85) \times 10^{-4}$			& --							& --										\\
O/H$^{(8)}$ 		& ($2.45 \pm 0.80) \times 10^{-4}$		& --    							& --							& --										\\
S/H$^{(9)}$ 		& ($4.45 \pm 1.05) \times 10^{-5}$		& --                             				& -- & -- \\%($3.24 \pm 1.62) \times 10^{-4}$	& ($3.24 \pm 1.62) \times 10^{-4}$ 				\\
P/H$^{(10)}$ 		& ($1.08 \pm 0.06) \times 10^{-6}$		& ($3.64 \pm 0.24) \times 10^{-6}$		& --							& --										\\			
\hline
\end{tabular}\\
$^{(1)}$ \citet{vonZahn1998} and \citet{Niemann1998} for Jupiter, \citet{Conrath2000} and \citet{Atreya2016} for Saturn, \citet{Conrath1987} for Uranus and \citet{Burgdorf2003} for Neptune. We only consider the higher value of the uncertainty on He in the case of Neptune. $^{(2-5)}$ \citet{Mahaffy2000} for Jupiter. $^{(6)}$ \citet{Wong2004} for Jupiter, \citet{Fletcher2009a} for Saturn, \citet{Lindal1987}, \citet{95baines}, \citet{Karkoschka2009}, and \citet{14sromovsky} for Uranus, \citet{Lindal1990}, \citet{95baines}, and \citet{11karkoschka} for Neptune. $^{(7)}$ \citet{Wong2004} for Jupiter, \citet{Fletcher2011} for Saturn (our N/H range is derived from the observed range of 90--500 ppm of NH$_3$). $^{(8)}$ \citet{Wong2004} for Jupiter (probably a lower limit, not representative of the bulk O/H). \citet{deGraauw1997} has detected H$_2$O at 5\,$\mu$m with ISO in Saturn, but the measurement at 1--3\,bars is not representative of the bulk O/H. $^{(9)}$ \citet{Wong2004} for Jupiter.%, \citet{Briggs1989} for Saturn.%, \citet{91depater} for Uranus and Neptune. 
$^{(10)}$ \citet{Fletcher2009b} for Jupiter and Saturn.
\end{center}
\end{table*}

\clearpage

\begin{table*}
\begin{center}
\caption[]{Ratios to protosolar values in the upper tropospheres of Jupiter, Saturn, Uranus and Neptune \label{table2}}
\begin{tabular}{lllll}
\hline
\noalign{\smallskip}
Elements	& Jupiter/Protosolar$^{(1)}$	& Saturn/Protosolar$^{(1)}$	&  Uranus/Protosolar$^{(1)}$	&  Neptune/Protosolar$^{(1)}$\\
\hline
\noalign{\smallskip}
He/H		& $0.81 \pm 0.05$			& $0.70 \pm 0.14$			& $0.93 \pm 0.21$			& $0.93 \pm 0.16$			\\
Ne/H		& $0.10 \pm 0.03$			& --						& --						& --						\\
Ar/H		& $2.55 \pm 0.83$			& --						& --						& --						\\
Kr/H		& $2.16 \pm 0.59$			& --						& --						& --						\\
Xe/H		& $2.12 \pm 0.59$			& --						& --						& --						\\
C/H		& $4.27 \pm 1.13$			& $9.61 \pm 0.59$			& $\sim$20 -- 120		        & $\sim$20 -- 120   	                 \\
N/H		& $4.06 \pm 2.02$			& 0.61 -- 3.48				& --						& --						\\
O/H		& $0.40 \pm 0.15$ (hotspot)	& --    					& --						& --						\\
S/H		& $2.73 \pm 0.65$			& --    					& -- & -- \\ %$20.78 \pm 10.40$			& $20.78 \pm 10.40$ 		\\
P/H		& $3.30 \pm 0.37$			& $11.17 \pm 1.31$			& --						& --						\\			
\hline
\end{tabular}\\
Error is defined as ($\Delta$E/E)$^2$ =  ($\Delta$X/X$_{\rm planet}$)$^2$ + ($\Delta$X/X$_{\rm Protosun}$)$^2$. $^{(1)}$ \citet{Lodders2009}. \\
\textit{Caveat:} These ratios only refer to the levels where abundance measurements have been performed, i.e. in the upper tropospheres. Thus, they are not automatically representative of deep interior enrichments. This is especially true if the deep interior contain a significant fraction of another element (e.g. oxygen in Uranus and Neptune, according to models). Moreover, the Helium value was computed for pure H$_2$/He mixtures (i.e. the upper tropospheric CH$_4$ has not been accounted for), because CH$_4$ is condensed at 1\,bar where He is measured. 
\end{center}
\end{table*}

\clearpage

\begin{table*}
\begin{center}
\caption[]{Isotopic ratios measured in Jupiter, Saturn, Uranus and Neptune \label{table3}}
\begin{tabular}{lcccccc}
\hline
\noalign{\smallskip}
Isotopic ratio							& Jupiter							& Saturn								&Uranus							& Neptune						\\
\hline
D/H (in H$_2$)$^{(1)}$					& (2.60 $\pm$ 0.7) $\times$ 10$^{-5}$	& $1.70 ^{+0.75}_{-0.45}$ $\times$ 10$^{-5}$	& (4.4 $\pm$ 0.4) $\times$ 10$^{-5}$		& (4.1 $\pm$ 0.4) $\times$ 10$^{-5}$		\\
$^3$He/$^4$He$^{(2)}$					& (1.66 $\pm$ 0.05) $\times$ 10$^{-4}$	& --									& --								& --  								\\
$^{12}$C/$^{13}$C (in CH$_4$)$^{(3)}$		& 92.6$^{+4.5}_{-4.1}$				& 91.8$^{+8.4}_{-7.8}$					& -- 								& -- 								\\
$^{14}$N/$^{15}$N (in NH$_3$)$^{(4)}$		& 434.8$^{+65}_{-50}$				& $>357$ 								& --								& --								\\
$^{20}$Ne/$^{22}$Ne$^{(5)}$				& 13 $\pm$ 2						& --									& --								& -- 								\\
$^{36}$Ar/$^{38}$Ar$^{(6)}$				& 5.6	 $\pm$ 0.25					& --									& --								& -- 								\\
$^{136}$Xe/total Xe$^{(7)}$				& 0.076 $\pm$	0.009				& --									& --								& --  								\\
$^{134}$Xe/total Xe$^{(8)}$				& 0.091 $\pm$ 0.007					& --									& --								& --  								\\
$^{132}$Xe/total Xe$^{(9)}$				& 0.290 $\pm$ 0.020					& --									& --								& -- 								\\
$^{131}$Xe/total Xe$^{(10)}$				& 0.203 $\pm$ 0.018					& --									& --								& --  								\\
$^{130}$Xe/total Xe$^{(11)}$				& 0.038 $\pm$ 0.005					& --									& --								& -- 								\\
$^{129}$Xe/total Xe$^{(12)}$				& 0.285 $\pm$ 0.021					& --									& --								& --  								\\
$^{128}$Xe/total Xe$^{(13)}$				& 0.018 $\pm$ 0.002					& --									& --								& -- 								\\
\hline
\end{tabular}\\
$^{(1)}$ \cite{Mahaffy1998} for Jupiter, \cite{Lellouch2001} for Saturn, \cite{Feuchtgruber2013} for Uranus and Neptune. $^{(2)}$ \cite{Mahaffy1998} for Jupiter. $^{(3)}$ \cite{Niemann1998} for Jupiter, \cite{Fletcher2009a} for Saturn. $^{(4)}$ \cite{Wong2004} for Jupiter, \cite{Fletcher2014} for Saturn. $^{(5-13)}$ \cite{Mahaffy2000} for Jupiter.
\end{center}
\end{table*}

\clearpage

\begin{table*}
\begin{center}
\caption[]{Measurement requirements}
\small{\begin{tabular}{ll}
\hline
\noalign{\smallskip}
Instrument					& Measurement									\\
\hline
Mass spectrometer				& Elemental and chemical composition					\\
							& Isotopic composition								\\
							& High molecular mass organics						\\
Helium Abundance Detector		& Helium abundance									\\							
Atmospheric Structure Instrument	& Pressure, temperature, density, molecular weight profile		\\
Doppler Wind Experiment			& Measure winds, speed and direction					\\							
Nephelometer					& Cloud structure									\\
							& Solid/liquid particles								\\
Net-flux radiometer				& Thermal/solar energy								\\									
\hline
\end{tabular}}
\end{center}
\label{payload_1}
\end{table*}

\clearpage

%\begin{table*}
%\begin{center}
%\caption[]{Uranus and Neptune probe NFR will build on the Saturn probe development efforts at NASA/GSFC and is designed to meet giant planets baseline net flux measurement goals.}
%\small{\begin{tabular}{lcc}
%\hline
%\noalign{\smallskip}
%Parameter				& Saturn probe NFR  		& ice-giant probe NFR							\\
%\hline
%Spectral range			& 0.25 to 50 $\mu$m	 		& 0.25 to 300 $\mu$m							\\
%Optics				& Winston cones			& Winston cones								\\
%Spectral channels		& 2						& 7											\\
%Field-Of-View			&\multicolumn{2}{c}{5$\deg$}												\\
%Viewing angles	±		&\multicolumn{2}{c}{$\pm$80$\deg$; $\pm$45$\deg$ and 0$\deg$ relative to nadir/zenith}			\\
%Detectors (uncooled)		&	Two single element thermopiles	 & Seven single element thermopiles 			\\
%Pixel count			&	3 (1 dark pixel)	& 9 (2 dark pixels)										\\
%Pixel size				&	0.25 mm square	& 0.25 mm square 									\\
%Mass				&	$\sim$2.4 kg &	$\sim$2.4 kg 											\\
%Basic power			&	$\sim$5 W &	$\sim$5.2 W 											\\
%Envelope				&	(11 $\times$ 14 $\times$ 28) cm$^3$	 & (11 $\times$ 14 $\times$ 21) cm$^3$		\\
%Data volume (90 mins)	&	297 kbits & 	670 kbits 												\\	
%Operating modes		&	100 ms integration	& 85 ms integration									\\
%Observation strategy	 	&	\multicolumn{2}{c}{Sequential rotation into five sky view angles}					\\
%\hline
%\end{tabular}}
%\end{center}
%\label{payload_2}
%\end{table*}

\clearpage

\begin{table*}
\begin{center}
\caption[]{Seven baseline NFR spectral filter channels and objectives, for maximizing science return from both Uranus and Neptune's atmospheres.}
\small{\begin{tabular}{lcc}
\hline
Ch$\#$		&	Wavelength ($\mu$m)	& Objectives													\\
\hline
1 			&	2.5--300				&	Deposition/loss of thermal radiation								\\
2			&	50--100				&	Ammonia humidity at $>$ 1 bar 								\\
3			&	14--35				&	Water vapor												\\
4			&	8.5--8.8				&	cloud opacity; implanted sulphur species (SO$_2$, H$_2$S, etc.)		\\
5			&	3.5--5.8				&	Water vapor and cloud structure								\\
6			&	0.6--3.5 				&	Solar deposition of methane absorption; cloud particles				\\
7			&	0.2--3.5				&	Total deposition of solar radiation and hot spot detection				\\
\hline
\end{tabular}}
\end{center}
\label{payload_3}
\end{table*}

\end{document}